\documentclass[12pt]{article}
\usepackage{epsfig,amsfonts,amssymb}
\usepackage{hyperref}
\usepackage{cite}
\usepackage{cite}

\def\ZZZ{{\hbox{ Z\kern-1.6mm Z}}}
\def\RRR{{\hbox{ R\kern-2.4mm R}}}
\def\CCC{{\hbox{ C\kern-2.0mm C}}}
\def\zzz{{\hbox{z\kern-1mm z}}}

\newcommand{\qeq}{{\hbox{=\kern-2.3mm ? \kern.5mm }}}
\renewcommand{\qeq}{=}

\newcommand{\be}{\begin{equation}}
\newcommand{\ee}{\end{equation}}
\newcommand{\ben}{\begin{eqnarray}\displaystyle}
\newcommand{\een}{\end{eqnarray}}

\newcommand{\refb}[1]{(\ref{#1})}

\def\one{{\hbox{ 1\kern-.8mm l}}}
\def\zero{{\hbox{ 0\kern-1.5mm 0}}}

\newcommand{\bea}[1]{\begin{eqnarray}\label{#1} }
\newcommand{\eea}{\end{eqnarray}}

\newcommand{\eqref}{\refb}

\usepackage{bm}
\usepackage[table]{xcolor}

\def\figa{
\def\JPicScale{0.3}
\ifx\JPicScale\undefined\def\JPicScale{1}\fi
\unitlength \JPicScale mm
\begin{picture}(225,90)(0,0)
\linethickness{0.3mm}
\qbezier(10,40)(10.28,41.64)(16.1,52.8)
\qbezier(16.1,52.8)(21.93,63.97)(30,65)
\qbezier(30,65)(38.66,62.03)(40.16,46.65)
\qbezier(40.16,46.65)(41.67,31.27)(50,30)
\qbezier(50,30)(59.31,33.64)(57.5,52.5)
\qbezier(57.5,52.5)(55.69,71.36)(65,75)
\qbezier(65,75)(73.53,74.42)(75.62,61.09)
\qbezier(75.62,61.09)(77.72,47.77)(85,40)
\qbezier(85,40)(89.36,36.6)(94.78,34.75)
\qbezier(94.78,34.75)(100.21,32.89)(105,35)
\qbezier(105,35)(112.51,40.15)(113.02,51.56)
\qbezier(113.02,51.56)(113.52,62.98)(120,70)
\qbezier(120,70)(125.29,74.29)(132.21,76.08)
\qbezier(132.21,76.08)(139.13,77.88)(145,75)
\qbezier(145,75)(155.1,66.28)(154.6,47.03)
\qbezier(154.6,47.03)(154.1,27.78)(165,25)
\qbezier(165,25)(176.45,27.54)(177.37,49.31)
\qbezier(177.37,49.31)(178.3,71.07)(190,75)
\qbezier(190,75)(197.79,73.48)(200.06,59.78)
\qbezier(200.06,59.78)(202.34,46.08)(210,45)
\qbezier(210,45)(219.82,48.52)(222.58,65.58)
\qbezier(222.58,65.58)(225.33,82.65)(225,85)
\linethickness{0.3mm}
\put(5,30){\line(1,0){215}}
\put(220,30){\vector(1,0){0.12}}
\linethickness{0.3mm}
\put(5,30){\line(0,1){60}}
\put(5,90){\vector(0,1){0.12}}

\put(7,82){$V$}

\put(230, 25){$\phi$}

\end{picture}

}

\def\figb{

\def\JPicScale{0.4}
\ifx\JPicScale\undefined\def\JPicScale{1}\fi
\unitlength \JPicScale mm
\begin{picture}(130,95)(0,0)
\linethickness{0.3mm}
\put(10,40){\line(1,0){20}}
\linethickness{0.3mm}
\multiput(30,40)(0.24,0.12){83}{\line(1,0){0.24}}
\linethickness{0.3mm}
\multiput(30,40)(0.24,-0.12){83}{\line(1,0){0.24}}
\linethickness{0.3mm}
\multiput(50,50)(0.12,0.12){167}{\line(1,0){0.12}}
\linethickness{0.3mm}
\put(50,50){\line(1,0){20}}
\linethickness{0.3mm}
\put(50,30){\line(1,0){20}}
\linethickness{0.3mm}
\multiput(50,30)(0.12,-0.12){167}{\line(1,0){0.12}}
\linethickness{0.3mm}
\multiput(70,30)(0.24,0.12){83}{\line(1,0){0.24}}
\linethickness{0.3mm}
\multiput(70,30)(0.24,-0.12){83}{\line(1,0){0.24}}
\linethickness{0.3mm}
\multiput(90,40)(0.12,0.12){167}{\line(1,0){0.12}}
\linethickness{0.3mm}
\put(90,40){\line(1,0){20}}
\linethickness{0.3mm}
\put(90,20){\line(1,0){20}}
\linethickness{0.3mm}
\multiput(90,20)(0.12,-0.12){167}{\line(1,0){0.12}}
\linethickness{0.3mm}
\multiput(110,20)(0.24,0.12){83}{\line(1,0){0.24}}
\linethickness{0.3mm}
\multiput(110,20)(0.24,-0.12){83}{\line(1,0){0.24}}
\linethickness{0.3mm}
\multiput(70,70)(0.24,0.12){83}{\line(1,0){0.24}}
\linethickness{0.3mm}
\put(70,70){\line(1,0){20}}
\linethickness{0.3mm}
\multiput(110,60)(0.24,0.12){83}{\line(1,0){0.24}}
\linethickness{0.3mm}
\put(110,60){\line(1,0){20}}
\linethickness{0.3mm}
\multiput(90,80)(0.24,0.12){83}{\line(1,0){0.24}}
\linethickness{0.3mm}
\put(90,80){\line(1,0){20}}
\linethickness{0.3mm}
\multiput(110,80)(0.24,0.12){83}{\line(1,0){0.24}}
\linethickness{0.3mm}
\put(110,80){\line(1,0){20}}
\linethickness{0.3mm}
\put(110,-5){\line(0,1){10}}
\linethickness{0.3mm}
\put(70,5){\line(0,1){10}}
\linethickness{0.3mm}
\put(70,45){\line(0,1){10}}
\linethickness{0.3mm}
\put(90,65){\line(0,1){10}}
\linethickness{0.3mm}
\put(110,85){\line(0,1){10}}
\linethickness{0.3mm}
\put(110,35){\line(0,1){10}}
\end{picture}

}

\def\figc{

\def\JPicScale{0.8}
\ifx\JPicScale\undefined\def\JPicScale{1}\fi
\unitlength \JPicScale mm
\begin{picture}(90,90)(0,0)
\linethickness{0.3mm}
\put(10,10){\line(1,0){80}}
\put(10,10){\line(0,1){80}}
\put(90,10){\line(0,1){80}}
\put(10,90){\line(1,0){80}}
\linethickness{0.3mm}
\multiput(20,90)(0.12,-0.12){250}{\line(1,0){0.12}}
\linethickness{0.3mm}
\multiput(50,60)(0.12,0.12){250}{\line(1,0){0.12}}
\linethickness{0.3mm}
\multiput(25,90)(0.12,-0.12){83}{\line(1,0){0.12}}
\linethickness{0.3mm}
\multiput(35,80)(0.12,0.12){83}{\line(1,0){0.12}}
\linethickness{0.3mm}
\multiput(55,90)(0.12,-0.12){83}{\line(1,0){0.12}}
\linethickness{0.3mm}
\multiput(65,80)(0.12,0.12){83}{\line(1,0){0.12}}
\put(45,55){\makebox(0,0)[cc]{A}}

\put(80,85){\makebox(0,0)[cc]{B}}

\put(15,85){\makebox(0,0)[cc]{C}}

\put(35,80){\makebox(0,0)[cc]{P}}

\put(60,75){\makebox(0,0)[cc]{Q}}

\end{picture}

}

\def\fige{

\def\JPicScale{0.8}
\ifx\JPicScale\undefined\def\JPicScale{1}\fi
\unitlength \JPicScale mm
\begin{picture}(220.06,80.06)(0,0)
\linethickness{16.4mm}
\put(75.33,47.25){\line(0,1){0.5}}
\multiput(75.31,48.25)(0.03,-0.5){1}{\line(0,-1){0.5}}
\multiput(75.25,48.74)(0.05,-0.49){1}{\line(0,-1){0.49}}
\multiput(75.18,49.23)(0.08,-0.49){1}{\line(0,-1){0.49}}
\multiput(75.07,49.72)(0.1,-0.49){1}{\line(0,-1){0.49}}
\multiput(74.94,50.2)(0.13,-0.48){1}{\line(0,-1){0.48}}
\multiput(74.79,50.67)(0.15,-0.47){1}{\line(0,-1){0.47}}
\multiput(74.61,51.14)(0.18,-0.46){1}{\line(0,-1){0.46}}
\multiput(74.41,51.59)(0.1,-0.23){2}{\line(0,-1){0.23}}
\multiput(74.18,52.03)(0.11,-0.22){2}{\line(0,-1){0.22}}
\multiput(73.93,52.46)(0.12,-0.22){2}{\line(0,-1){0.22}}
\multiput(73.66,52.88)(0.14,-0.21){2}{\line(0,-1){0.21}}
\multiput(73.37,53.28)(0.15,-0.2){2}{\line(0,-1){0.2}}
\multiput(73.06,53.67)(0.1,-0.13){3}{\line(0,-1){0.13}}
\multiput(72.73,54.04)(0.11,-0.12){3}{\line(0,-1){0.12}}
\multiput(72.37,54.39)(0.12,-0.12){3}{\line(1,0){0.12}}
\multiput(72,54.72)(0.12,-0.11){3}{\line(1,0){0.12}}
\multiput(71.62,55.04)(0.13,-0.1){3}{\line(1,0){0.13}}
\multiput(71.22,55.33)(0.2,-0.15){2}{\line(1,0){0.2}}
\multiput(70.8,55.6)(0.21,-0.14){2}{\line(1,0){0.21}}
\multiput(70.37,55.85)(0.22,-0.12){2}{\line(1,0){0.22}}
\multiput(69.92,56.08)(0.22,-0.11){2}{\line(1,0){0.22}}
\multiput(69.47,56.28)(0.23,-0.1){2}{\line(1,0){0.23}}
\multiput(69,56.46)(0.46,-0.18){1}{\line(1,0){0.46}}
\multiput(68.53,56.61)(0.47,-0.15){1}{\line(1,0){0.47}}
\multiput(68.05,56.74)(0.48,-0.13){1}{\line(1,0){0.48}}
\multiput(67.56,56.84)(0.49,-0.1){1}{\line(1,0){0.49}}
\multiput(67.07,56.92)(0.49,-0.08){1}{\line(1,0){0.49}}
\multiput(66.58,56.97)(0.49,-0.05){1}{\line(1,0){0.49}}
\multiput(66.08,57)(0.5,-0.03){1}{\line(1,0){0.5}}
\put(65.58,57){\line(1,0){0.5}}
\multiput(65.09,56.97)(0.5,0.03){1}{\line(1,0){0.5}}
\multiput(64.59,56.92)(0.49,0.05){1}{\line(1,0){0.49}}
\multiput(64.1,56.84)(0.49,0.08){1}{\line(1,0){0.49}}
\multiput(63.62,56.74)(0.49,0.1){1}{\line(1,0){0.49}}
\multiput(63.13,56.61)(0.48,0.13){1}{\line(1,0){0.48}}
\multiput(62.66,56.46)(0.47,0.15){1}{\line(1,0){0.47}}
\multiput(62.2,56.28)(0.46,0.18){1}{\line(1,0){0.46}}
\multiput(61.74,56.08)(0.23,0.1){2}{\line(1,0){0.23}}
\multiput(61.3,55.85)(0.22,0.11){2}{\line(1,0){0.22}}
\multiput(60.87,55.6)(0.22,0.12){2}{\line(1,0){0.22}}
\multiput(60.45,55.33)(0.21,0.14){2}{\line(1,0){0.21}}
\multiput(60.05,55.04)(0.2,0.15){2}{\line(1,0){0.2}}
\multiput(59.66,54.72)(0.13,0.1){3}{\line(1,0){0.13}}
\multiput(59.29,54.39)(0.12,0.11){3}{\line(1,0){0.12}}
\multiput(58.94,54.04)(0.12,0.12){3}{\line(0,1){0.12}}
\multiput(58.61,53.67)(0.11,0.12){3}{\line(0,1){0.12}}
\multiput(58.3,53.28)(0.1,0.13){3}{\line(0,1){0.13}}
\multiput(58,52.88)(0.15,0.2){2}{\line(0,1){0.2}}
\multiput(57.73,52.46)(0.14,0.21){2}{\line(0,1){0.21}}
\multiput(57.48,52.03)(0.12,0.22){2}{\line(0,1){0.22}}
\multiput(57.26,51.59)(0.11,0.22){2}{\line(0,1){0.22}}
\multiput(57.06,51.14)(0.1,0.23){2}{\line(0,1){0.23}}
\multiput(56.88,50.67)(0.18,0.46){1}{\line(0,1){0.46}}
\multiput(56.72,50.2)(0.15,0.47){1}{\line(0,1){0.47}}
\multiput(56.59,49.72)(0.13,0.48){1}{\line(0,1){0.48}}
\multiput(56.49,49.23)(0.1,0.49){1}{\line(0,1){0.49}}
\multiput(56.41,48.74)(0.08,0.49){1}{\line(0,1){0.49}}
\multiput(56.36,48.25)(0.05,0.49){1}{\line(0,1){0.49}}
\multiput(56.34,47.75)(0.03,0.5){1}{\line(0,1){0.5}}
\put(56.34,47.25){\line(0,1){0.5}}
\multiput(56.34,47.25)(0.03,-0.5){1}{\line(0,-1){0.5}}
\multiput(56.36,46.75)(0.05,-0.49){1}{\line(0,-1){0.49}}
\multiput(56.41,46.26)(0.08,-0.49){1}{\line(0,-1){0.49}}
\multiput(56.49,45.77)(0.1,-0.49){1}{\line(0,-1){0.49}}
\multiput(56.59,45.28)(0.13,-0.48){1}{\line(0,-1){0.48}}
\multiput(56.72,44.8)(0.15,-0.47){1}{\line(0,-1){0.47}}
\multiput(56.88,44.33)(0.18,-0.46){1}{\line(0,-1){0.46}}
\multiput(57.06,43.86)(0.1,-0.23){2}{\line(0,-1){0.23}}
\multiput(57.26,43.41)(0.11,-0.22){2}{\line(0,-1){0.22}}
\multiput(57.48,42.97)(0.12,-0.22){2}{\line(0,-1){0.22}}
\multiput(57.73,42.54)(0.14,-0.21){2}{\line(0,-1){0.21}}
\multiput(58,42.12)(0.15,-0.2){2}{\line(0,-1){0.2}}
\multiput(58.3,41.72)(0.1,-0.13){3}{\line(0,-1){0.13}}
\multiput(58.61,41.33)(0.11,-0.12){3}{\line(0,-1){0.12}}
\multiput(58.94,40.96)(0.12,-0.12){3}{\line(1,0){0.12}}
\multiput(59.29,40.61)(0.12,-0.11){3}{\line(1,0){0.12}}
\multiput(59.66,40.28)(0.13,-0.1){3}{\line(1,0){0.13}}
\multiput(60.05,39.96)(0.2,-0.15){2}{\line(1,0){0.2}}
\multiput(60.45,39.67)(0.21,-0.14){2}{\line(1,0){0.21}}
\multiput(60.87,39.4)(0.22,-0.12){2}{\line(1,0){0.22}}
\multiput(61.3,39.15)(0.22,-0.11){2}{\line(1,0){0.22}}
\multiput(61.74,38.92)(0.23,-0.1){2}{\line(1,0){0.23}}
\multiput(62.2,38.72)(0.46,-0.18){1}{\line(1,0){0.46}}
\multiput(62.66,38.54)(0.47,-0.15){1}{\line(1,0){0.47}}
\multiput(63.13,38.39)(0.48,-0.13){1}{\line(1,0){0.48}}
\multiput(63.62,38.26)(0.49,-0.1){1}{\line(1,0){0.49}}
\multiput(64.1,38.16)(0.49,-0.08){1}{\line(1,0){0.49}}
\multiput(64.59,38.08)(0.49,-0.05){1}{\line(1,0){0.49}}
\multiput(65.09,38.03)(0.5,-0.03){1}{\line(1,0){0.5}}
\put(65.58,38){\line(1,0){0.5}}
\multiput(66.08,38)(0.5,0.03){1}{\line(1,0){0.5}}
\multiput(66.58,38.03)(0.49,0.05){1}{\line(1,0){0.49}}
\multiput(67.07,38.08)(0.49,0.08){1}{\line(1,0){0.49}}
\multiput(67.56,38.16)(0.49,0.1){1}{\line(1,0){0.49}}
\multiput(68.05,38.26)(0.48,0.13){1}{\line(1,0){0.48}}
\multiput(68.53,38.39)(0.47,0.15){1}{\line(1,0){0.47}}
\multiput(69,38.54)(0.46,0.18){1}{\line(1,0){0.46}}
\multiput(69.47,38.72)(0.23,0.1){2}{\line(1,0){0.23}}
\multiput(69.92,38.92)(0.22,0.11){2}{\line(1,0){0.22}}
\multiput(70.37,39.15)(0.22,0.12){2}{\line(1,0){0.22}}
\multiput(70.8,39.4)(0.21,0.14){2}{\line(1,0){0.21}}
\multiput(71.22,39.67)(0.2,0.15){2}{\line(1,0){0.2}}
\multiput(71.62,39.96)(0.13,0.1){3}{\line(1,0){0.13}}
\multiput(72,40.28)(0.12,0.11){3}{\line(1,0){0.12}}
\multiput(72.37,40.61)(0.12,0.12){3}{\line(1,0){0.12}}
\multiput(72.73,40.96)(0.11,0.12){3}{\line(0,1){0.12}}
\multiput(73.06,41.33)(0.1,0.13){3}{\line(0,1){0.13}}
\multiput(73.37,41.72)(0.15,0.2){2}{\line(0,1){0.2}}
\multiput(73.66,42.12)(0.14,0.21){2}{\line(0,1){0.21}}
\multiput(73.93,42.54)(0.12,0.22){2}{\line(0,1){0.22}}
\multiput(74.18,42.97)(0.11,0.22){2}{\line(0,1){0.22}}
\multiput(74.41,43.41)(0.1,0.23){2}{\line(0,1){0.23}}
\multiput(74.61,43.86)(0.18,0.46){1}{\line(0,1){0.46}}
\multiput(74.79,44.33)(0.15,0.47){1}{\line(0,1){0.47}}
\multiput(74.94,44.8)(0.13,0.48){1}{\line(0,1){0.48}}
\multiput(75.07,45.28)(0.1,0.49){1}{\line(0,1){0.49}}
\multiput(75.18,45.77)(0.08,0.49){1}{\line(0,1){0.49}}
\multiput(75.25,46.26)(0.05,0.49){1}{\line(0,1){0.49}}
\multiput(75.31,46.75)(0.03,0.5){1}{\line(0,1){0.5}}

\linethickness{15.05mm}
\put(120.4,67.25){\line(0,1){0.5}}
\multiput(120.37,68.24)(0.03,-0.5){1}{\line(0,-1){0.5}}
\multiput(120.31,68.74)(0.06,-0.49){1}{\line(0,-1){0.49}}
\multiput(120.22,69.22)(0.09,-0.49){1}{\line(0,-1){0.49}}
\multiput(120.09,69.71)(0.12,-0.48){1}{\line(0,-1){0.48}}
\multiput(119.94,70.18)(0.15,-0.47){1}{\line(0,-1){0.47}}
\multiput(119.76,70.64)(0.09,-0.23){2}{\line(0,-1){0.23}}
\multiput(119.54,71.09)(0.11,-0.22){2}{\line(0,-1){0.22}}
\multiput(119.3,71.52)(0.12,-0.22){2}{\line(0,-1){0.22}}
\multiput(119.04,71.94)(0.13,-0.21){2}{\line(0,-1){0.21}}
\multiput(118.75,72.35)(0.15,-0.2){2}{\line(0,-1){0.2}}
\multiput(118.43,72.73)(0.11,-0.13){3}{\line(0,-1){0.13}}
\multiput(118.09,73.09)(0.11,-0.12){3}{\line(0,-1){0.12}}
\multiput(117.73,73.43)(0.12,-0.11){3}{\line(1,0){0.12}}
\multiput(117.35,73.75)(0.13,-0.11){3}{\line(1,0){0.13}}
\multiput(116.94,74.04)(0.2,-0.15){2}{\line(1,0){0.2}}
\multiput(116.52,74.3)(0.21,-0.13){2}{\line(1,0){0.21}}
\multiput(116.09,74.54)(0.22,-0.12){2}{\line(1,0){0.22}}
\multiput(115.64,74.76)(0.22,-0.11){2}{\line(1,0){0.22}}
\multiput(115.18,74.94)(0.23,-0.09){2}{\line(1,0){0.23}}
\multiput(114.71,75.09)(0.47,-0.15){1}{\line(1,0){0.47}}
\multiput(114.22,75.22)(0.48,-0.12){1}{\line(1,0){0.48}}
\multiput(113.74,75.31)(0.49,-0.09){1}{\line(1,0){0.49}}
\multiput(113.24,75.37)(0.49,-0.06){1}{\line(1,0){0.49}}
\multiput(112.75,75.4)(0.5,-0.03){1}{\line(1,0){0.5}}
\put(112.25,75.4){\line(1,0){0.5}}
\multiput(111.76,75.37)(0.5,0.03){1}{\line(1,0){0.5}}
\multiput(111.26,75.31)(0.49,0.06){1}{\line(1,0){0.49}}
\multiput(110.78,75.22)(0.49,0.09){1}{\line(1,0){0.49}}
\multiput(110.29,75.09)(0.48,0.12){1}{\line(1,0){0.48}}
\multiput(109.82,74.94)(0.47,0.15){1}{\line(1,0){0.47}}
\multiput(109.36,74.76)(0.23,0.09){2}{\line(1,0){0.23}}
\multiput(108.91,74.54)(0.22,0.11){2}{\line(1,0){0.22}}
\multiput(108.48,74.3)(0.22,0.12){2}{\line(1,0){0.22}}
\multiput(108.06,74.04)(0.21,0.13){2}{\line(1,0){0.21}}
\multiput(107.65,73.75)(0.2,0.15){2}{\line(1,0){0.2}}
\multiput(107.27,73.43)(0.13,0.11){3}{\line(1,0){0.13}}
\multiput(106.91,73.09)(0.12,0.11){3}{\line(1,0){0.12}}
\multiput(106.57,72.73)(0.11,0.12){3}{\line(0,1){0.12}}
\multiput(106.25,72.35)(0.11,0.13){3}{\line(0,1){0.13}}
\multiput(105.96,71.94)(0.15,0.2){2}{\line(0,1){0.2}}
\multiput(105.7,71.52)(0.13,0.21){2}{\line(0,1){0.21}}
\multiput(105.46,71.09)(0.12,0.22){2}{\line(0,1){0.22}}
\multiput(105.24,70.64)(0.11,0.22){2}{\line(0,1){0.22}}
\multiput(105.06,70.18)(0.09,0.23){2}{\line(0,1){0.23}}
\multiput(104.91,69.71)(0.15,0.47){1}{\line(0,1){0.47}}
\multiput(104.78,69.22)(0.12,0.48){1}{\line(0,1){0.48}}
\multiput(104.69,68.74)(0.09,0.49){1}{\line(0,1){0.49}}
\multiput(104.63,68.24)(0.06,0.49){1}{\line(0,1){0.49}}
\multiput(104.6,67.75)(0.03,0.5){1}{\line(0,1){0.5}}
\put(104.6,67.25){\line(0,1){0.5}}
\multiput(104.6,67.25)(0.03,-0.5){1}{\line(0,-1){0.5}}
\multiput(104.63,66.76)(0.06,-0.49){1}{\line(0,-1){0.49}}
\multiput(104.69,66.26)(0.09,-0.49){1}{\line(0,-1){0.49}}
\multiput(104.78,65.78)(0.12,-0.48){1}{\line(0,-1){0.48}}
\multiput(104.91,65.29)(0.15,-0.47){1}{\line(0,-1){0.47}}
\multiput(105.06,64.82)(0.09,-0.23){2}{\line(0,-1){0.23}}
\multiput(105.24,64.36)(0.11,-0.22){2}{\line(0,-1){0.22}}
\multiput(105.46,63.91)(0.12,-0.22){2}{\line(0,-1){0.22}}
\multiput(105.7,63.48)(0.13,-0.21){2}{\line(0,-1){0.21}}
\multiput(105.96,63.06)(0.15,-0.2){2}{\line(0,-1){0.2}}
\multiput(106.25,62.65)(0.11,-0.13){3}{\line(0,-1){0.13}}
\multiput(106.57,62.27)(0.11,-0.12){3}{\line(0,-1){0.12}}
\multiput(106.91,61.91)(0.12,-0.11){3}{\line(1,0){0.12}}
\multiput(107.27,61.57)(0.13,-0.11){3}{\line(1,0){0.13}}
\multiput(107.65,61.25)(0.2,-0.15){2}{\line(1,0){0.2}}
\multiput(108.06,60.96)(0.21,-0.13){2}{\line(1,0){0.21}}
\multiput(108.48,60.7)(0.22,-0.12){2}{\line(1,0){0.22}}
\multiput(108.91,60.46)(0.22,-0.11){2}{\line(1,0){0.22}}
\multiput(109.36,60.24)(0.23,-0.09){2}{\line(1,0){0.23}}
\multiput(109.82,60.06)(0.47,-0.15){1}{\line(1,0){0.47}}
\multiput(110.29,59.91)(0.48,-0.12){1}{\line(1,0){0.48}}
\multiput(110.78,59.78)(0.49,-0.09){1}{\line(1,0){0.49}}
\multiput(111.26,59.69)(0.49,-0.06){1}{\line(1,0){0.49}}
\multiput(111.76,59.63)(0.5,-0.03){1}{\line(1,0){0.5}}
\put(112.25,59.6){\line(1,0){0.5}}
\multiput(112.75,59.6)(0.5,0.03){1}{\line(1,0){0.5}}
\multiput(113.24,59.63)(0.49,0.06){1}{\line(1,0){0.49}}
\multiput(113.74,59.69)(0.49,0.09){1}{\line(1,0){0.49}}
\multiput(114.22,59.78)(0.48,0.12){1}{\line(1,0){0.48}}
\multiput(114.71,59.91)(0.47,0.15){1}{\line(1,0){0.47}}
\multiput(115.18,60.06)(0.23,0.09){2}{\line(1,0){0.23}}
\multiput(115.64,60.24)(0.22,0.11){2}{\line(1,0){0.22}}
\multiput(116.09,60.46)(0.22,0.12){2}{\line(1,0){0.22}}
\multiput(116.52,60.7)(0.21,0.13){2}{\line(1,0){0.21}}
\multiput(116.94,60.96)(0.2,0.15){2}{\line(1,0){0.2}}
\multiput(117.35,61.25)(0.13,0.11){3}{\line(1,0){0.13}}
\multiput(117.73,61.57)(0.12,0.11){3}{\line(1,0){0.12}}
\multiput(118.09,61.91)(0.11,0.12){3}{\line(0,1){0.12}}
\multiput(118.43,62.27)(0.11,0.13){3}{\line(0,1){0.13}}
\multiput(118.75,62.65)(0.15,0.2){2}{\line(0,1){0.2}}
\multiput(119.04,63.06)(0.13,0.21){2}{\line(0,1){0.21}}
\multiput(119.3,63.48)(0.12,0.22){2}{\line(0,1){0.22}}
\multiput(119.54,63.91)(0.11,0.22){2}{\line(0,1){0.22}}
\multiput(119.76,64.36)(0.09,0.23){2}{\line(0,1){0.23}}
\multiput(119.94,64.82)(0.15,0.47){1}{\line(0,1){0.47}}
\multiput(120.09,65.29)(0.12,0.48){1}{\line(0,1){0.48}}
\multiput(120.22,65.78)(0.09,0.49){1}{\line(0,1){0.49}}
\multiput(120.31,66.26)(0.06,0.49){1}{\line(0,1){0.49}}
\multiput(120.37,66.76)(0.03,0.5){1}{\line(0,1){0.5}}

\linethickness{15.05mm}
\put(175.4,42.25){\line(0,1){0.5}}
\multiput(175.37,43.24)(0.03,-0.5){1}{\line(0,-1){0.5}}
\multiput(175.31,43.74)(0.06,-0.49){1}{\line(0,-1){0.49}}
\multiput(175.22,44.22)(0.09,-0.49){1}{\line(0,-1){0.49}}
\multiput(175.09,44.71)(0.12,-0.48){1}{\line(0,-1){0.48}}
\multiput(174.94,45.18)(0.15,-0.47){1}{\line(0,-1){0.47}}
\multiput(174.76,45.64)(0.09,-0.23){2}{\line(0,-1){0.23}}
\multiput(174.54,46.09)(0.11,-0.22){2}{\line(0,-1){0.22}}
\multiput(174.3,46.52)(0.12,-0.22){2}{\line(0,-1){0.22}}
\multiput(174.04,46.94)(0.13,-0.21){2}{\line(0,-1){0.21}}
\multiput(173.75,47.35)(0.15,-0.2){2}{\line(0,-1){0.2}}
\multiput(173.43,47.73)(0.11,-0.13){3}{\line(0,-1){0.13}}
\multiput(173.09,48.09)(0.11,-0.12){3}{\line(0,-1){0.12}}
\multiput(172.73,48.43)(0.12,-0.11){3}{\line(1,0){0.12}}
\multiput(172.35,48.75)(0.13,-0.11){3}{\line(1,0){0.13}}
\multiput(171.94,49.04)(0.2,-0.15){2}{\line(1,0){0.2}}
\multiput(171.52,49.3)(0.21,-0.13){2}{\line(1,0){0.21}}
\multiput(171.09,49.54)(0.22,-0.12){2}{\line(1,0){0.22}}
\multiput(170.64,49.76)(0.22,-0.11){2}{\line(1,0){0.22}}
\multiput(170.18,49.94)(0.23,-0.09){2}{\line(1,0){0.23}}
\multiput(169.71,50.09)(0.47,-0.15){1}{\line(1,0){0.47}}
\multiput(169.22,50.22)(0.48,-0.12){1}{\line(1,0){0.48}}
\multiput(168.74,50.31)(0.49,-0.09){1}{\line(1,0){0.49}}
\multiput(168.24,50.37)(0.49,-0.06){1}{\line(1,0){0.49}}
\multiput(167.75,50.4)(0.5,-0.03){1}{\line(1,0){0.5}}
\put(167.25,50.4){\line(1,0){0.5}}
\multiput(166.76,50.37)(0.5,0.03){1}{\line(1,0){0.5}}
\multiput(166.26,50.31)(0.49,0.06){1}{\line(1,0){0.49}}
\multiput(165.78,50.22)(0.49,0.09){1}{\line(1,0){0.49}}
\multiput(165.29,50.09)(0.48,0.12){1}{\line(1,0){0.48}}
\multiput(164.82,49.94)(0.47,0.15){1}{\line(1,0){0.47}}
\multiput(164.36,49.76)(0.23,0.09){2}{\line(1,0){0.23}}
\multiput(163.91,49.54)(0.22,0.11){2}{\line(1,0){0.22}}
\multiput(163.48,49.3)(0.22,0.12){2}{\line(1,0){0.22}}
\multiput(163.06,49.04)(0.21,0.13){2}{\line(1,0){0.21}}
\multiput(162.65,48.75)(0.2,0.15){2}{\line(1,0){0.2}}
\multiput(162.27,48.43)(0.13,0.11){3}{\line(1,0){0.13}}
\multiput(161.91,48.09)(0.12,0.11){3}{\line(1,0){0.12}}
\multiput(161.57,47.73)(0.11,0.12){3}{\line(0,1){0.12}}
\multiput(161.25,47.35)(0.11,0.13){3}{\line(0,1){0.13}}
\multiput(160.96,46.94)(0.15,0.2){2}{\line(0,1){0.2}}
\multiput(160.7,46.52)(0.13,0.21){2}{\line(0,1){0.21}}
\multiput(160.46,46.09)(0.12,0.22){2}{\line(0,1){0.22}}
\multiput(160.24,45.64)(0.11,0.22){2}{\line(0,1){0.22}}
\multiput(160.06,45.18)(0.09,0.23){2}{\line(0,1){0.23}}
\multiput(159.91,44.71)(0.15,0.47){1}{\line(0,1){0.47}}
\multiput(159.78,44.22)(0.12,0.48){1}{\line(0,1){0.48}}
\multiput(159.69,43.74)(0.09,0.49){1}{\line(0,1){0.49}}
\multiput(159.63,43.24)(0.06,0.49){1}{\line(0,1){0.49}}
\multiput(159.6,42.75)(0.03,0.5){1}{\line(0,1){0.5}}
\put(159.6,42.25){\line(0,1){0.5}}
\multiput(159.6,42.25)(0.03,-0.5){1}{\line(0,-1){0.5}}
\multiput(159.63,41.76)(0.06,-0.49){1}{\line(0,-1){0.49}}
\multiput(159.69,41.26)(0.09,-0.49){1}{\line(0,-1){0.49}}
\multiput(159.78,40.78)(0.12,-0.48){1}{\line(0,-1){0.48}}
\multiput(159.91,40.29)(0.15,-0.47){1}{\line(0,-1){0.47}}
\multiput(160.06,39.82)(0.09,-0.23){2}{\line(0,-1){0.23}}
\multiput(160.24,39.36)(0.11,-0.22){2}{\line(0,-1){0.22}}
\multiput(160.46,38.91)(0.12,-0.22){2}{\line(0,-1){0.22}}
\multiput(160.7,38.48)(0.13,-0.21){2}{\line(0,-1){0.21}}
\multiput(160.96,38.06)(0.15,-0.2){2}{\line(0,-1){0.2}}
\multiput(161.25,37.65)(0.11,-0.13){3}{\line(0,-1){0.13}}
\multiput(161.57,37.27)(0.11,-0.12){3}{\line(0,-1){0.12}}
\multiput(161.91,36.91)(0.12,-0.11){3}{\line(1,0){0.12}}
\multiput(162.27,36.57)(0.13,-0.11){3}{\line(1,0){0.13}}
\multiput(162.65,36.25)(0.2,-0.15){2}{\line(1,0){0.2}}
\multiput(163.06,35.96)(0.21,-0.13){2}{\line(1,0){0.21}}
\multiput(163.48,35.7)(0.22,-0.12){2}{\line(1,0){0.22}}
\multiput(163.91,35.46)(0.22,-0.11){2}{\line(1,0){0.22}}
\multiput(164.36,35.24)(0.23,-0.09){2}{\line(1,0){0.23}}
\multiput(164.82,35.06)(0.47,-0.15){1}{\line(1,0){0.47}}
\multiput(165.29,34.91)(0.48,-0.12){1}{\line(1,0){0.48}}
\multiput(165.78,34.78)(0.49,-0.09){1}{\line(1,0){0.49}}
\multiput(166.26,34.69)(0.49,-0.06){1}{\line(1,0){0.49}}
\multiput(166.76,34.63)(0.5,-0.03){1}{\line(1,0){0.5}}
\put(167.25,34.6){\line(1,0){0.5}}
\multiput(167.75,34.6)(0.5,0.03){1}{\line(1,0){0.5}}
\multiput(168.24,34.63)(0.49,0.06){1}{\line(1,0){0.49}}
\multiput(168.74,34.69)(0.49,0.09){1}{\line(1,0){0.49}}
\multiput(169.22,34.78)(0.48,0.12){1}{\line(1,0){0.48}}
\multiput(169.71,34.91)(0.47,0.15){1}{\line(1,0){0.47}}
\multiput(170.18,35.06)(0.23,0.09){2}{\line(1,0){0.23}}
\multiput(170.64,35.24)(0.22,0.11){2}{\line(1,0){0.22}}
\multiput(171.09,35.46)(0.22,0.12){2}{\line(1,0){0.22}}
\multiput(171.52,35.7)(0.21,0.13){2}{\line(1,0){0.21}}
\multiput(171.94,35.96)(0.2,0.15){2}{\line(1,0){0.2}}
\multiput(172.35,36.25)(0.13,0.11){3}{\line(1,0){0.13}}
\multiput(172.73,36.57)(0.12,0.11){3}{\line(1,0){0.12}}
\multiput(173.09,36.91)(0.11,0.12){3}{\line(0,1){0.12}}
\multiput(173.43,37.27)(0.11,0.13){3}{\line(0,1){0.13}}
\multiput(173.75,37.65)(0.15,0.2){2}{\line(0,1){0.2}}
\multiput(174.04,38.06)(0.13,0.21){2}{\line(0,1){0.21}}
\multiput(174.3,38.48)(0.12,0.22){2}{\line(0,1){0.22}}
\multiput(174.54,38.91)(0.11,0.22){2}{\line(0,1){0.22}}
\multiput(174.76,39.36)(0.09,0.23){2}{\line(0,1){0.23}}
\multiput(174.94,39.82)(0.15,0.47){1}{\line(0,1){0.47}}
\multiput(175.09,40.29)(0.12,0.48){1}{\line(0,1){0.48}}
\multiput(175.22,40.78)(0.09,0.49){1}{\line(0,1){0.49}}
\multiput(175.31,41.26)(0.06,0.49){1}{\line(0,1){0.49}}
\multiput(175.37,41.76)(0.03,0.5){1}{\line(0,1){0.5}}

\linethickness{15.05mm}
\put(214.17,74.17){\circle{11.79}}

\end{picture}

}

\def\figd{

\def\JPicScale{0.5}
\ifx\JPicScale\undefined\def\JPicScale{1}\fi
\unitlength \JPicScale mm
\begin{picture}(215.2,72.07)(0,0)
\linethickness{0.3mm}
\put(40.03,15.58){\line(0,1){0.5}}
\multiput(40.01,16.59)(0.02,-0.5){1}{\line(0,-1){0.5}}
\multiput(39.96,17.09)(0.05,-0.5){1}{\line(0,-1){0.5}}
\multiput(39.89,17.58)(0.07,-0.5){1}{\line(0,-1){0.5}}
\multiput(39.8,18.08)(0.09,-0.49){1}{\line(0,-1){0.49}}
\multiput(39.68,18.56)(0.11,-0.49){1}{\line(0,-1){0.49}}
\multiput(39.55,19.05)(0.14,-0.48){1}{\line(0,-1){0.48}}
\multiput(39.39,19.52)(0.16,-0.48){1}{\line(0,-1){0.48}}
\multiput(39.2,19.99)(0.09,-0.23){2}{\line(0,-1){0.23}}
\multiput(39,20.45)(0.1,-0.23){2}{\line(0,-1){0.23}}
\multiput(38.78,20.9)(0.11,-0.22){2}{\line(0,-1){0.22}}
\multiput(38.53,21.34)(0.12,-0.22){2}{\line(0,-1){0.22}}
\multiput(38.27,21.77)(0.13,-0.21){2}{\line(0,-1){0.21}}
\multiput(37.99,22.18)(0.14,-0.21){2}{\line(0,-1){0.21}}
\multiput(37.68,22.58)(0.1,-0.13){3}{\line(0,-1){0.13}}
\multiput(37.36,22.97)(0.11,-0.13){3}{\line(0,-1){0.13}}
\multiput(37.03,23.34)(0.11,-0.12){3}{\line(0,-1){0.12}}
\multiput(36.67,23.69)(0.12,-0.12){3}{\line(1,0){0.12}}
\multiput(36.3,24.03)(0.12,-0.11){3}{\line(1,0){0.12}}
\multiput(35.91,24.35)(0.13,-0.11){3}{\line(1,0){0.13}}
\multiput(35.51,24.65)(0.13,-0.1){3}{\line(1,0){0.13}}
\multiput(35.1,24.94)(0.21,-0.14){2}{\line(1,0){0.21}}
\multiput(34.67,25.2)(0.21,-0.13){2}{\line(1,0){0.21}}
\multiput(34.23,25.45)(0.22,-0.12){2}{\line(1,0){0.22}}
\multiput(33.78,25.67)(0.22,-0.11){2}{\line(1,0){0.22}}
\multiput(33.32,25.87)(0.23,-0.1){2}{\line(1,0){0.23}}
\multiput(32.86,26.05)(0.23,-0.09){2}{\line(1,0){0.23}}
\multiput(32.38,26.21)(0.48,-0.16){1}{\line(1,0){0.48}}
\multiput(31.9,26.35)(0.48,-0.14){1}{\line(1,0){0.48}}
\multiput(31.41,26.46)(0.49,-0.11){1}{\line(1,0){0.49}}
\multiput(30.92,26.56)(0.49,-0.09){1}{\line(1,0){0.49}}
\multiput(30.42,26.63)(0.5,-0.07){1}{\line(1,0){0.5}}
\multiput(29.92,26.67)(0.5,-0.05){1}{\line(1,0){0.5}}
\multiput(29.42,26.7)(0.5,-0.02){1}{\line(1,0){0.5}}
\put(28.92,26.7){\line(1,0){0.5}}
\multiput(28.41,26.67)(0.5,0.02){1}{\line(1,0){0.5}}
\multiput(27.91,26.63)(0.5,0.05){1}{\line(1,0){0.5}}
\multiput(27.42,26.56)(0.5,0.07){1}{\line(1,0){0.5}}
\multiput(26.92,26.46)(0.49,0.09){1}{\line(1,0){0.49}}
\multiput(26.44,26.35)(0.49,0.11){1}{\line(1,0){0.49}}
\multiput(25.95,26.21)(0.48,0.14){1}{\line(1,0){0.48}}
\multiput(25.48,26.05)(0.48,0.16){1}{\line(1,0){0.48}}
\multiput(25.01,25.87)(0.23,0.09){2}{\line(1,0){0.23}}
\multiput(24.55,25.67)(0.23,0.1){2}{\line(1,0){0.23}}
\multiput(24.1,25.45)(0.22,0.11){2}{\line(1,0){0.22}}
\multiput(23.66,25.2)(0.22,0.12){2}{\line(1,0){0.22}}
\multiput(23.23,24.94)(0.21,0.13){2}{\line(1,0){0.21}}
\multiput(22.82,24.65)(0.21,0.14){2}{\line(1,0){0.21}}
\multiput(22.42,24.35)(0.13,0.1){3}{\line(1,0){0.13}}
\multiput(22.03,24.03)(0.13,0.11){3}{\line(1,0){0.13}}
\multiput(21.66,23.69)(0.12,0.11){3}{\line(1,0){0.12}}
\multiput(21.31,23.34)(0.12,0.12){3}{\line(1,0){0.12}}
\multiput(20.97,22.97)(0.11,0.12){3}{\line(0,1){0.12}}
\multiput(20.65,22.58)(0.11,0.13){3}{\line(0,1){0.13}}
\multiput(20.35,22.18)(0.1,0.13){3}{\line(0,1){0.13}}
\multiput(20.06,21.77)(0.14,0.21){2}{\line(0,1){0.21}}
\multiput(19.8,21.34)(0.13,0.21){2}{\line(0,1){0.21}}
\multiput(19.55,20.9)(0.12,0.22){2}{\line(0,1){0.22}}
\multiput(19.33,20.45)(0.11,0.22){2}{\line(0,1){0.22}}
\multiput(19.13,19.99)(0.1,0.23){2}{\line(0,1){0.23}}
\multiput(18.95,19.52)(0.09,0.23){2}{\line(0,1){0.23}}
\multiput(18.79,19.05)(0.16,0.48){1}{\line(0,1){0.48}}
\multiput(18.65,18.56)(0.14,0.48){1}{\line(0,1){0.48}}
\multiput(18.54,18.08)(0.11,0.49){1}{\line(0,1){0.49}}
\multiput(18.44,17.58)(0.09,0.49){1}{\line(0,1){0.49}}
\multiput(18.37,17.09)(0.07,0.5){1}{\line(0,1){0.5}}
\multiput(18.33,16.59)(0.05,0.5){1}{\line(0,1){0.5}}
\multiput(18.3,16.08)(0.02,0.5){1}{\line(0,1){0.5}}
\put(18.3,15.58){\line(0,1){0.5}}
\multiput(18.3,15.58)(0.02,-0.5){1}{\line(0,-1){0.5}}
\multiput(18.33,15.08)(0.05,-0.5){1}{\line(0,-1){0.5}}
\multiput(18.37,14.58)(0.07,-0.5){1}{\line(0,-1){0.5}}
\multiput(18.44,14.08)(0.09,-0.49){1}{\line(0,-1){0.49}}
\multiput(18.54,13.59)(0.11,-0.49){1}{\line(0,-1){0.49}}
\multiput(18.65,13.1)(0.14,-0.48){1}{\line(0,-1){0.48}}
\multiput(18.79,12.62)(0.16,-0.48){1}{\line(0,-1){0.48}}
\multiput(18.95,12.14)(0.09,-0.23){2}{\line(0,-1){0.23}}
\multiput(19.13,11.68)(0.1,-0.23){2}{\line(0,-1){0.23}}
\multiput(19.33,11.22)(0.11,-0.22){2}{\line(0,-1){0.22}}
\multiput(19.55,10.77)(0.12,-0.22){2}{\line(0,-1){0.22}}
\multiput(19.8,10.33)(0.13,-0.21){2}{\line(0,-1){0.21}}
\multiput(20.06,9.9)(0.14,-0.21){2}{\line(0,-1){0.21}}
\multiput(20.35,9.49)(0.1,-0.13){3}{\line(0,-1){0.13}}
\multiput(20.65,9.09)(0.11,-0.13){3}{\line(0,-1){0.13}}
\multiput(20.97,8.7)(0.11,-0.12){3}{\line(0,-1){0.12}}
\multiput(21.31,8.33)(0.12,-0.12){3}{\line(1,0){0.12}}
\multiput(21.66,7.97)(0.12,-0.11){3}{\line(1,0){0.12}}
\multiput(22.03,7.64)(0.13,-0.11){3}{\line(1,0){0.13}}
\multiput(22.42,7.32)(0.13,-0.1){3}{\line(1,0){0.13}}
\multiput(22.82,7.01)(0.21,-0.14){2}{\line(1,0){0.21}}
\multiput(23.23,6.73)(0.21,-0.13){2}{\line(1,0){0.21}}
\multiput(23.66,6.47)(0.22,-0.12){2}{\line(1,0){0.22}}
\multiput(24.1,6.22)(0.22,-0.11){2}{\line(1,0){0.22}}
\multiput(24.55,6)(0.23,-0.1){2}{\line(1,0){0.23}}
\multiput(25.01,5.8)(0.23,-0.09){2}{\line(1,0){0.23}}
\multiput(25.48,5.61)(0.48,-0.16){1}{\line(1,0){0.48}}
\multiput(25.95,5.45)(0.48,-0.14){1}{\line(1,0){0.48}}
\multiput(26.44,5.32)(0.49,-0.11){1}{\line(1,0){0.49}}
\multiput(26.92,5.2)(0.49,-0.09){1}{\line(1,0){0.49}}
\multiput(27.42,5.11)(0.5,-0.07){1}{\line(1,0){0.5}}
\multiput(27.91,5.04)(0.5,-0.05){1}{\line(1,0){0.5}}
\multiput(28.41,4.99)(0.5,-0.02){1}{\line(1,0){0.5}}
\put(28.92,4.97){\line(1,0){0.5}}
\multiput(29.42,4.97)(0.5,0.02){1}{\line(1,0){0.5}}
\multiput(29.92,4.99)(0.5,0.05){1}{\line(1,0){0.5}}
\multiput(30.42,5.04)(0.5,0.07){1}{\line(1,0){0.5}}
\multiput(30.92,5.11)(0.49,0.09){1}{\line(1,0){0.49}}
\multiput(31.41,5.2)(0.49,0.11){1}{\line(1,0){0.49}}
\multiput(31.9,5.32)(0.48,0.14){1}{\line(1,0){0.48}}
\multiput(32.38,5.45)(0.48,0.16){1}{\line(1,0){0.48}}
\multiput(32.86,5.61)(0.23,0.09){2}{\line(1,0){0.23}}
\multiput(33.32,5.8)(0.23,0.1){2}{\line(1,0){0.23}}
\multiput(33.78,6)(0.22,0.11){2}{\line(1,0){0.22}}
\multiput(34.23,6.22)(0.22,0.12){2}{\line(1,0){0.22}}
\multiput(34.67,6.47)(0.21,0.13){2}{\line(1,0){0.21}}
\multiput(35.1,6.73)(0.21,0.14){2}{\line(1,0){0.21}}
\multiput(35.51,7.01)(0.13,0.1){3}{\line(1,0){0.13}}
\multiput(35.91,7.32)(0.13,0.11){3}{\line(1,0){0.13}}
\multiput(36.3,7.64)(0.12,0.11){3}{\line(1,0){0.12}}
\multiput(36.67,7.97)(0.12,0.12){3}{\line(0,1){0.12}}
\multiput(37.03,8.33)(0.11,0.12){3}{\line(0,1){0.12}}
\multiput(37.36,8.7)(0.11,0.13){3}{\line(0,1){0.13}}
\multiput(37.68,9.09)(0.1,0.13){3}{\line(0,1){0.13}}
\multiput(37.99,9.49)(0.14,0.21){2}{\line(0,1){0.21}}
\multiput(38.27,9.9)(0.13,0.21){2}{\line(0,1){0.21}}
\multiput(38.53,10.33)(0.12,0.22){2}{\line(0,1){0.22}}
\multiput(38.78,10.77)(0.11,0.22){2}{\line(0,1){0.22}}
\multiput(39,11.22)(0.1,0.23){2}{\line(0,1){0.23}}
\multiput(39.2,11.68)(0.09,0.23){2}{\line(0,1){0.23}}
\multiput(39.39,12.14)(0.16,0.48){1}{\line(0,1){0.48}}
\multiput(39.55,12.62)(0.14,0.48){1}{\line(0,1){0.48}}
\multiput(39.68,13.1)(0.11,0.49){1}{\line(0,1){0.49}}
\multiput(39.8,13.59)(0.09,0.49){1}{\line(0,1){0.49}}
\multiput(39.89,14.08)(0.07,0.5){1}{\line(0,1){0.5}}
\multiput(39.96,14.58)(0.05,0.5){1}{\line(0,1){0.5}}
\multiput(40.01,15.08)(0.02,0.5){1}{\line(0,1){0.5}}

\linethickness{0.3mm}
\put(62.07,64.75){\line(0,1){0.5}}
\multiput(62.03,65.76)(0.04,-0.5){1}{\line(0,-1){0.5}}
\multiput(61.96,66.26)(0.07,-0.5){1}{\line(0,-1){0.5}}
\multiput(61.85,66.75)(0.11,-0.49){1}{\line(0,-1){0.49}}
\multiput(61.71,67.23)(0.14,-0.48){1}{\line(0,-1){0.48}}
\multiput(61.53,67.71)(0.18,-0.47){1}{\line(0,-1){0.47}}
\multiput(61.32,68.17)(0.1,-0.23){2}{\line(0,-1){0.23}}
\multiput(61.08,68.61)(0.12,-0.22){2}{\line(0,-1){0.22}}
\multiput(60.81,69.03)(0.14,-0.21){2}{\line(0,-1){0.21}}
\multiput(60.51,69.44)(0.1,-0.13){3}{\line(0,-1){0.13}}
\multiput(60.18,69.82)(0.11,-0.13){3}{\line(0,-1){0.13}}
\multiput(59.82,70.18)(0.12,-0.12){3}{\line(1,0){0.12}}
\multiput(59.44,70.51)(0.13,-0.11){3}{\line(1,0){0.13}}
\multiput(59.03,70.81)(0.13,-0.1){3}{\line(1,0){0.13}}
\multiput(58.61,71.08)(0.21,-0.14){2}{\line(1,0){0.21}}
\multiput(58.17,71.32)(0.22,-0.12){2}{\line(1,0){0.22}}
\multiput(57.71,71.53)(0.23,-0.1){2}{\line(1,0){0.23}}
\multiput(57.23,71.71)(0.47,-0.18){1}{\line(1,0){0.47}}
\multiput(56.75,71.85)(0.48,-0.14){1}{\line(1,0){0.48}}
\multiput(56.26,71.96)(0.49,-0.11){1}{\line(1,0){0.49}}
\multiput(55.76,72.03)(0.5,-0.07){1}{\line(1,0){0.5}}
\multiput(55.25,72.07)(0.5,-0.04){1}{\line(1,0){0.5}}
\put(54.75,72.07){\line(1,0){0.5}}
\multiput(54.24,72.03)(0.5,0.04){1}{\line(1,0){0.5}}
\multiput(53.74,71.96)(0.5,0.07){1}{\line(1,0){0.5}}
\multiput(53.25,71.85)(0.49,0.11){1}{\line(1,0){0.49}}
\multiput(52.77,71.71)(0.48,0.14){1}{\line(1,0){0.48}}
\multiput(52.29,71.53)(0.47,0.18){1}{\line(1,0){0.47}}
\multiput(51.83,71.32)(0.23,0.1){2}{\line(1,0){0.23}}
\multiput(51.39,71.08)(0.22,0.12){2}{\line(1,0){0.22}}
\multiput(50.97,70.81)(0.21,0.14){2}{\line(1,0){0.21}}
\multiput(50.56,70.51)(0.13,0.1){3}{\line(1,0){0.13}}
\multiput(50.18,70.18)(0.13,0.11){3}{\line(1,0){0.13}}
\multiput(49.82,69.82)(0.12,0.12){3}{\line(1,0){0.12}}
\multiput(49.49,69.44)(0.11,0.13){3}{\line(0,1){0.13}}
\multiput(49.19,69.03)(0.1,0.13){3}{\line(0,1){0.13}}
\multiput(48.92,68.61)(0.14,0.21){2}{\line(0,1){0.21}}
\multiput(48.68,68.17)(0.12,0.22){2}{\line(0,1){0.22}}
\multiput(48.47,67.71)(0.1,0.23){2}{\line(0,1){0.23}}
\multiput(48.29,67.23)(0.18,0.47){1}{\line(0,1){0.47}}
\multiput(48.15,66.75)(0.14,0.48){1}{\line(0,1){0.48}}
\multiput(48.04,66.26)(0.11,0.49){1}{\line(0,1){0.49}}
\multiput(47.97,65.76)(0.07,0.5){1}{\line(0,1){0.5}}
\multiput(47.93,65.25)(0.04,0.5){1}{\line(0,1){0.5}}
\put(47.93,64.75){\line(0,1){0.5}}
\multiput(47.93,64.75)(0.04,-0.5){1}{\line(0,-1){0.5}}
\multiput(47.97,64.24)(0.07,-0.5){1}{\line(0,-1){0.5}}
\multiput(48.04,63.74)(0.11,-0.49){1}{\line(0,-1){0.49}}
\multiput(48.15,63.25)(0.14,-0.48){1}{\line(0,-1){0.48}}
\multiput(48.29,62.77)(0.18,-0.47){1}{\line(0,-1){0.47}}
\multiput(48.47,62.29)(0.1,-0.23){2}{\line(0,-1){0.23}}
\multiput(48.68,61.83)(0.12,-0.22){2}{\line(0,-1){0.22}}
\multiput(48.92,61.39)(0.14,-0.21){2}{\line(0,-1){0.21}}
\multiput(49.19,60.97)(0.1,-0.13){3}{\line(0,-1){0.13}}
\multiput(49.49,60.56)(0.11,-0.13){3}{\line(0,-1){0.13}}
\multiput(49.82,60.18)(0.12,-0.12){3}{\line(1,0){0.12}}
\multiput(50.18,59.82)(0.13,-0.11){3}{\line(1,0){0.13}}
\multiput(50.56,59.49)(0.13,-0.1){3}{\line(1,0){0.13}}
\multiput(50.97,59.19)(0.21,-0.14){2}{\line(1,0){0.21}}
\multiput(51.39,58.92)(0.22,-0.12){2}{\line(1,0){0.22}}
\multiput(51.83,58.68)(0.23,-0.1){2}{\line(1,0){0.23}}
\multiput(52.29,58.47)(0.47,-0.18){1}{\line(1,0){0.47}}
\multiput(52.77,58.29)(0.48,-0.14){1}{\line(1,0){0.48}}
\multiput(53.25,58.15)(0.49,-0.11){1}{\line(1,0){0.49}}
\multiput(53.74,58.04)(0.5,-0.07){1}{\line(1,0){0.5}}
\multiput(54.24,57.97)(0.5,-0.04){1}{\line(1,0){0.5}}
\put(54.75,57.93){\line(1,0){0.5}}
\multiput(55.25,57.93)(0.5,0.04){1}{\line(1,0){0.5}}
\multiput(55.76,57.97)(0.5,0.07){1}{\line(1,0){0.5}}
\multiput(56.26,58.04)(0.49,0.11){1}{\line(1,0){0.49}}
\multiput(56.75,58.15)(0.48,0.14){1}{\line(1,0){0.48}}
\multiput(57.23,58.29)(0.47,0.18){1}{\line(1,0){0.47}}
\multiput(57.71,58.47)(0.23,0.1){2}{\line(1,0){0.23}}
\multiput(58.17,58.68)(0.22,0.12){2}{\line(1,0){0.22}}
\multiput(58.61,58.92)(0.21,0.14){2}{\line(1,0){0.21}}
\multiput(59.03,59.19)(0.13,0.1){3}{\line(1,0){0.13}}
\multiput(59.44,59.49)(0.13,0.11){3}{\line(1,0){0.13}}
\multiput(59.82,59.82)(0.12,0.12){3}{\line(1,0){0.12}}
\multiput(60.18,60.18)(0.11,0.13){3}{\line(0,1){0.13}}
\multiput(60.51,60.56)(0.1,0.13){3}{\line(0,1){0.13}}
\multiput(60.81,60.97)(0.14,0.21){2}{\line(0,1){0.21}}
\multiput(61.08,61.39)(0.12,0.22){2}{\line(0,1){0.22}}
\multiput(61.32,61.83)(0.1,0.23){2}{\line(0,1){0.23}}
\multiput(61.53,62.29)(0.18,0.47){1}{\line(0,1){0.47}}
\multiput(61.71,62.77)(0.14,0.48){1}{\line(0,1){0.48}}
\multiput(61.85,63.25)(0.11,0.49){1}{\line(0,1){0.49}}
\multiput(61.96,63.74)(0.07,0.5){1}{\line(0,1){0.5}}
\multiput(62.03,64.24)(0.04,0.5){1}{\line(0,1){0.5}}

\linethickness{0.3mm}
\put(128.51,39.75){\line(0,1){0.5}}
\multiput(128.49,40.75)(0.02,-0.5){1}{\line(0,-1){0.5}}
\multiput(128.46,41.26)(0.03,-0.5){1}{\line(0,-1){0.5}}
\multiput(128.41,41.76)(0.05,-0.5){1}{\line(0,-1){0.5}}
\multiput(128.35,42.26)(0.06,-0.5){1}{\line(0,-1){0.5}}
\multiput(128.27,42.75)(0.08,-0.5){1}{\line(0,-1){0.5}}
\multiput(128.18,43.25)(0.09,-0.49){1}{\line(0,-1){0.49}}
\multiput(128.07,43.74)(0.11,-0.49){1}{\line(0,-1){0.49}}
\multiput(127.94,44.22)(0.13,-0.49){1}{\line(0,-1){0.49}}
\multiput(127.8,44.71)(0.14,-0.48){1}{\line(0,-1){0.48}}
\multiput(127.64,45.19)(0.16,-0.48){1}{\line(0,-1){0.48}}
\multiput(127.47,45.66)(0.17,-0.47){1}{\line(0,-1){0.47}}
\multiput(127.29,46.13)(0.09,-0.23){2}{\line(0,-1){0.23}}
\multiput(127.09,46.59)(0.1,-0.23){2}{\line(0,-1){0.23}}
\multiput(126.88,47.04)(0.11,-0.23){2}{\line(0,-1){0.23}}
\multiput(126.65,47.49)(0.11,-0.22){2}{\line(0,-1){0.22}}
\multiput(126.4,47.93)(0.12,-0.22){2}{\line(0,-1){0.22}}
\multiput(126.15,48.36)(0.13,-0.22){2}{\line(0,-1){0.22}}
\multiput(125.88,48.79)(0.13,-0.21){2}{\line(0,-1){0.21}}
\multiput(125.6,49.2)(0.14,-0.21){2}{\line(0,-1){0.21}}
\multiput(125.3,49.61)(0.15,-0.2){2}{\line(0,-1){0.2}}
\multiput(124.99,50.01)(0.1,-0.13){3}{\line(0,-1){0.13}}
\multiput(124.67,50.4)(0.11,-0.13){3}{\line(0,-1){0.13}}
\multiput(124.34,50.77)(0.11,-0.13){3}{\line(0,-1){0.13}}
\multiput(124,51.14)(0.11,-0.12){3}{\line(0,-1){0.12}}
\multiput(123.64,51.5)(0.12,-0.12){3}{\line(1,0){0.12}}
\multiput(123.27,51.84)(0.12,-0.11){3}{\line(1,0){0.12}}
\multiput(122.9,52.17)(0.13,-0.11){3}{\line(1,0){0.13}}
\multiput(122.51,52.49)(0.13,-0.11){3}{\line(1,0){0.13}}
\multiput(122.11,52.8)(0.13,-0.1){3}{\line(1,0){0.13}}
\multiput(121.7,53.1)(0.2,-0.15){2}{\line(1,0){0.2}}
\multiput(121.29,53.38)(0.21,-0.14){2}{\line(1,0){0.21}}
\multiput(120.86,53.65)(0.21,-0.13){2}{\line(1,0){0.21}}
\multiput(120.43,53.9)(0.22,-0.13){2}{\line(1,0){0.22}}
\multiput(119.99,54.15)(0.22,-0.12){2}{\line(1,0){0.22}}
\multiput(119.54,54.38)(0.22,-0.11){2}{\line(1,0){0.22}}
\multiput(119.09,54.59)(0.23,-0.11){2}{\line(1,0){0.23}}
\multiput(118.63,54.79)(0.23,-0.1){2}{\line(1,0){0.23}}
\multiput(118.16,54.97)(0.23,-0.09){2}{\line(1,0){0.23}}
\multiput(117.69,55.14)(0.47,-0.17){1}{\line(1,0){0.47}}
\multiput(117.21,55.3)(0.48,-0.16){1}{\line(1,0){0.48}}
\multiput(116.72,55.44)(0.48,-0.14){1}{\line(1,0){0.48}}
\multiput(116.24,55.57)(0.49,-0.13){1}{\line(1,0){0.49}}
\multiput(115.75,55.68)(0.49,-0.11){1}{\line(1,0){0.49}}
\multiput(115.25,55.77)(0.49,-0.09){1}{\line(1,0){0.49}}
\multiput(114.76,55.85)(0.5,-0.08){1}{\line(1,0){0.5}}
\multiput(114.26,55.91)(0.5,-0.06){1}{\line(1,0){0.5}}
\multiput(113.76,55.96)(0.5,-0.05){1}{\line(1,0){0.5}}
\multiput(113.25,55.99)(0.5,-0.03){1}{\line(1,0){0.5}}
\multiput(112.75,56.01)(0.5,-0.02){1}{\line(1,0){0.5}}
\put(112.25,56.01){\line(1,0){0.5}}
\multiput(111.75,55.99)(0.5,0.02){1}{\line(1,0){0.5}}
\multiput(111.24,55.96)(0.5,0.03){1}{\line(1,0){0.5}}
\multiput(110.74,55.91)(0.5,0.05){1}{\line(1,0){0.5}}
\multiput(110.24,55.85)(0.5,0.06){1}{\line(1,0){0.5}}
\multiput(109.75,55.77)(0.5,0.08){1}{\line(1,0){0.5}}
\multiput(109.25,55.68)(0.49,0.09){1}{\line(1,0){0.49}}
\multiput(108.76,55.57)(0.49,0.11){1}{\line(1,0){0.49}}
\multiput(108.28,55.44)(0.49,0.13){1}{\line(1,0){0.49}}
\multiput(107.79,55.3)(0.48,0.14){1}{\line(1,0){0.48}}
\multiput(107.31,55.14)(0.48,0.16){1}{\line(1,0){0.48}}
\multiput(106.84,54.97)(0.47,0.17){1}{\line(1,0){0.47}}
\multiput(106.37,54.79)(0.23,0.09){2}{\line(1,0){0.23}}
\multiput(105.91,54.59)(0.23,0.1){2}{\line(1,0){0.23}}
\multiput(105.46,54.38)(0.23,0.11){2}{\line(1,0){0.23}}
\multiput(105.01,54.15)(0.22,0.11){2}{\line(1,0){0.22}}
\multiput(104.57,53.9)(0.22,0.12){2}{\line(1,0){0.22}}
\multiput(104.14,53.65)(0.22,0.13){2}{\line(1,0){0.22}}
\multiput(103.71,53.38)(0.21,0.13){2}{\line(1,0){0.21}}
\multiput(103.3,53.1)(0.21,0.14){2}{\line(1,0){0.21}}
\multiput(102.89,52.8)(0.2,0.15){2}{\line(1,0){0.2}}
\multiput(102.49,52.49)(0.13,0.1){3}{\line(1,0){0.13}}
\multiput(102.1,52.17)(0.13,0.11){3}{\line(1,0){0.13}}
\multiput(101.73,51.84)(0.13,0.11){3}{\line(1,0){0.13}}
\multiput(101.36,51.5)(0.12,0.11){3}{\line(1,0){0.12}}
\multiput(101,51.14)(0.12,0.12){3}{\line(1,0){0.12}}
\multiput(100.66,50.77)(0.11,0.12){3}{\line(0,1){0.12}}
\multiput(100.33,50.4)(0.11,0.13){3}{\line(0,1){0.13}}
\multiput(100.01,50.01)(0.11,0.13){3}{\line(0,1){0.13}}
\multiput(99.7,49.61)(0.1,0.13){3}{\line(0,1){0.13}}
\multiput(99.4,49.2)(0.15,0.2){2}{\line(0,1){0.2}}
\multiput(99.12,48.79)(0.14,0.21){2}{\line(0,1){0.21}}
\multiput(98.85,48.36)(0.13,0.21){2}{\line(0,1){0.21}}
\multiput(98.6,47.93)(0.13,0.22){2}{\line(0,1){0.22}}
\multiput(98.35,47.49)(0.12,0.22){2}{\line(0,1){0.22}}
\multiput(98.12,47.04)(0.11,0.22){2}{\line(0,1){0.22}}
\multiput(97.91,46.59)(0.11,0.23){2}{\line(0,1){0.23}}
\multiput(97.71,46.13)(0.1,0.23){2}{\line(0,1){0.23}}
\multiput(97.53,45.66)(0.09,0.23){2}{\line(0,1){0.23}}
\multiput(97.36,45.19)(0.17,0.47){1}{\line(0,1){0.47}}
\multiput(97.2,44.71)(0.16,0.48){1}{\line(0,1){0.48}}
\multiput(97.06,44.22)(0.14,0.48){1}{\line(0,1){0.48}}
\multiput(96.93,43.74)(0.13,0.49){1}{\line(0,1){0.49}}
\multiput(96.82,43.25)(0.11,0.49){1}{\line(0,1){0.49}}
\multiput(96.73,42.75)(0.09,0.49){1}{\line(0,1){0.49}}
\multiput(96.65,42.26)(0.08,0.5){1}{\line(0,1){0.5}}
\multiput(96.59,41.76)(0.06,0.5){1}{\line(0,1){0.5}}
\multiput(96.54,41.26)(0.05,0.5){1}{\line(0,1){0.5}}
\multiput(96.51,40.75)(0.03,0.5){1}{\line(0,1){0.5}}
\multiput(96.49,40.25)(0.02,0.5){1}{\line(0,1){0.5}}
\put(96.49,39.75){\line(0,1){0.5}}
\multiput(96.49,39.75)(0.02,-0.5){1}{\line(0,-1){0.5}}
\multiput(96.51,39.25)(0.03,-0.5){1}{\line(0,-1){0.5}}
\multiput(96.54,38.74)(0.05,-0.5){1}{\line(0,-1){0.5}}
\multiput(96.59,38.24)(0.06,-0.5){1}{\line(0,-1){0.5}}
\multiput(96.65,37.74)(0.08,-0.5){1}{\line(0,-1){0.5}}
\multiput(96.73,37.25)(0.09,-0.49){1}{\line(0,-1){0.49}}
\multiput(96.82,36.75)(0.11,-0.49){1}{\line(0,-1){0.49}}
\multiput(96.93,36.26)(0.13,-0.49){1}{\line(0,-1){0.49}}
\multiput(97.06,35.78)(0.14,-0.48){1}{\line(0,-1){0.48}}
\multiput(97.2,35.29)(0.16,-0.48){1}{\line(0,-1){0.48}}
\multiput(97.36,34.81)(0.17,-0.47){1}{\line(0,-1){0.47}}
\multiput(97.53,34.34)(0.09,-0.23){2}{\line(0,-1){0.23}}
\multiput(97.71,33.87)(0.1,-0.23){2}{\line(0,-1){0.23}}
\multiput(97.91,33.41)(0.11,-0.23){2}{\line(0,-1){0.23}}
\multiput(98.12,32.96)(0.11,-0.22){2}{\line(0,-1){0.22}}
\multiput(98.35,32.51)(0.12,-0.22){2}{\line(0,-1){0.22}}
\multiput(98.6,32.07)(0.13,-0.22){2}{\line(0,-1){0.22}}
\multiput(98.85,31.64)(0.13,-0.21){2}{\line(0,-1){0.21}}
\multiput(99.12,31.21)(0.14,-0.21){2}{\line(0,-1){0.21}}
\multiput(99.4,30.8)(0.15,-0.2){2}{\line(0,-1){0.2}}
\multiput(99.7,30.39)(0.1,-0.13){3}{\line(0,-1){0.13}}
\multiput(100.01,29.99)(0.11,-0.13){3}{\line(0,-1){0.13}}
\multiput(100.33,29.6)(0.11,-0.13){3}{\line(0,-1){0.13}}
\multiput(100.66,29.23)(0.11,-0.12){3}{\line(0,-1){0.12}}
\multiput(101,28.86)(0.12,-0.12){3}{\line(1,0){0.12}}
\multiput(101.36,28.5)(0.12,-0.11){3}{\line(1,0){0.12}}
\multiput(101.73,28.16)(0.13,-0.11){3}{\line(1,0){0.13}}
\multiput(102.1,27.83)(0.13,-0.11){3}{\line(1,0){0.13}}
\multiput(102.49,27.51)(0.13,-0.1){3}{\line(1,0){0.13}}
\multiput(102.89,27.2)(0.2,-0.15){2}{\line(1,0){0.2}}
\multiput(103.3,26.9)(0.21,-0.14){2}{\line(1,0){0.21}}
\multiput(103.71,26.62)(0.21,-0.13){2}{\line(1,0){0.21}}
\multiput(104.14,26.35)(0.22,-0.13){2}{\line(1,0){0.22}}
\multiput(104.57,26.1)(0.22,-0.12){2}{\line(1,0){0.22}}
\multiput(105.01,25.85)(0.22,-0.11){2}{\line(1,0){0.22}}
\multiput(105.46,25.62)(0.23,-0.11){2}{\line(1,0){0.23}}
\multiput(105.91,25.41)(0.23,-0.1){2}{\line(1,0){0.23}}
\multiput(106.37,25.21)(0.23,-0.09){2}{\line(1,0){0.23}}
\multiput(106.84,25.03)(0.47,-0.17){1}{\line(1,0){0.47}}
\multiput(107.31,24.86)(0.48,-0.16){1}{\line(1,0){0.48}}
\multiput(107.79,24.7)(0.48,-0.14){1}{\line(1,0){0.48}}
\multiput(108.28,24.56)(0.49,-0.13){1}{\line(1,0){0.49}}
\multiput(108.76,24.43)(0.49,-0.11){1}{\line(1,0){0.49}}
\multiput(109.25,24.32)(0.49,-0.09){1}{\line(1,0){0.49}}
\multiput(109.75,24.23)(0.5,-0.08){1}{\line(1,0){0.5}}
\multiput(110.24,24.15)(0.5,-0.06){1}{\line(1,0){0.5}}
\multiput(110.74,24.09)(0.5,-0.05){1}{\line(1,0){0.5}}
\multiput(111.24,24.04)(0.5,-0.03){1}{\line(1,0){0.5}}
\multiput(111.75,24.01)(0.5,-0.02){1}{\line(1,0){0.5}}
\put(112.25,23.99){\line(1,0){0.5}}
\multiput(112.75,23.99)(0.5,0.02){1}{\line(1,0){0.5}}
\multiput(113.25,24.01)(0.5,0.03){1}{\line(1,0){0.5}}
\multiput(113.76,24.04)(0.5,0.05){1}{\line(1,0){0.5}}
\multiput(114.26,24.09)(0.5,0.06){1}{\line(1,0){0.5}}
\multiput(114.76,24.15)(0.5,0.08){1}{\line(1,0){0.5}}
\multiput(115.25,24.23)(0.49,0.09){1}{\line(1,0){0.49}}
\multiput(115.75,24.32)(0.49,0.11){1}{\line(1,0){0.49}}
\multiput(116.24,24.43)(0.49,0.13){1}{\line(1,0){0.49}}
\multiput(116.72,24.56)(0.48,0.14){1}{\line(1,0){0.48}}
\multiput(117.21,24.7)(0.48,0.16){1}{\line(1,0){0.48}}
\multiput(117.69,24.86)(0.47,0.17){1}{\line(1,0){0.47}}
\multiput(118.16,25.03)(0.23,0.09){2}{\line(1,0){0.23}}
\multiput(118.63,25.21)(0.23,0.1){2}{\line(1,0){0.23}}
\multiput(119.09,25.41)(0.23,0.11){2}{\line(1,0){0.23}}
\multiput(119.54,25.62)(0.22,0.11){2}{\line(1,0){0.22}}
\multiput(119.99,25.85)(0.22,0.12){2}{\line(1,0){0.22}}
\multiput(120.43,26.1)(0.22,0.13){2}{\line(1,0){0.22}}
\multiput(120.86,26.35)(0.21,0.13){2}{\line(1,0){0.21}}
\multiput(121.29,26.62)(0.21,0.14){2}{\line(1,0){0.21}}
\multiput(121.7,26.9)(0.2,0.15){2}{\line(1,0){0.2}}
\multiput(122.11,27.2)(0.13,0.1){3}{\line(1,0){0.13}}
\multiput(122.51,27.51)(0.13,0.11){3}{\line(1,0){0.13}}
\multiput(122.9,27.83)(0.13,0.11){3}{\line(1,0){0.13}}
\multiput(123.27,28.16)(0.12,0.11){3}{\line(1,0){0.12}}
\multiput(123.64,28.5)(0.12,0.12){3}{\line(1,0){0.12}}
\multiput(124,28.86)(0.11,0.12){3}{\line(0,1){0.12}}
\multiput(124.34,29.23)(0.11,0.13){3}{\line(0,1){0.13}}
\multiput(124.67,29.6)(0.11,0.13){3}{\line(0,1){0.13}}
\multiput(124.99,29.99)(0.1,0.13){3}{\line(0,1){0.13}}
\multiput(125.3,30.39)(0.15,0.2){2}{\line(0,1){0.2}}
\multiput(125.6,30.8)(0.14,0.21){2}{\line(0,1){0.21}}
\multiput(125.88,31.21)(0.13,0.21){2}{\line(0,1){0.21}}
\multiput(126.15,31.64)(0.13,0.22){2}{\line(0,1){0.22}}
\multiput(126.4,32.07)(0.12,0.22){2}{\line(0,1){0.22}}
\multiput(126.65,32.51)(0.11,0.22){2}{\line(0,1){0.22}}
\multiput(126.88,32.96)(0.11,0.23){2}{\line(0,1){0.23}}
\multiput(127.09,33.41)(0.1,0.23){2}{\line(0,1){0.23}}
\multiput(127.29,33.87)(0.09,0.23){2}{\line(0,1){0.23}}
\multiput(127.47,34.34)(0.17,0.47){1}{\line(0,1){0.47}}
\multiput(127.64,34.81)(0.16,0.48){1}{\line(0,1){0.48}}
\multiput(127.8,35.29)(0.14,0.48){1}{\line(0,1){0.48}}
\multiput(127.94,35.78)(0.13,0.49){1}{\line(0,1){0.49}}
\multiput(128.07,36.26)(0.11,0.49){1}{\line(0,1){0.49}}
\multiput(128.18,36.75)(0.09,0.49){1}{\line(0,1){0.49}}
\multiput(128.27,37.25)(0.08,0.5){1}{\line(0,1){0.5}}
\multiput(128.35,37.74)(0.06,0.5){1}{\line(0,1){0.5}}
\multiput(128.41,38.24)(0.05,0.5){1}{\line(0,1){0.5}}
\multiput(128.46,38.74)(0.03,0.5){1}{\line(0,1){0.5}}
\multiput(128.49,39.25)(0.02,0.5){1}{\line(0,1){0.5}}

\linethickness{0.3mm}
\put(172.5,67.5){\circle{7.07}}

\linethickness{0.3mm}
\put(215.19,32.25){\line(0,1){0.5}}
\multiput(215.18,33.26)(0.02,-0.5){1}{\line(0,-1){0.5}}
\multiput(215.15,33.76)(0.03,-0.5){1}{\line(0,-1){0.5}}
\multiput(215.1,34.26)(0.05,-0.5){1}{\line(0,-1){0.5}}
\multiput(215.04,34.76)(0.06,-0.5){1}{\line(0,-1){0.5}}
\multiput(214.96,35.26)(0.08,-0.5){1}{\line(0,-1){0.5}}
\multiput(214.86,35.75)(0.09,-0.49){1}{\line(0,-1){0.49}}
\multiput(214.75,36.24)(0.11,-0.49){1}{\line(0,-1){0.49}}
\multiput(214.63,36.73)(0.13,-0.49){1}{\line(0,-1){0.49}}
\multiput(214.49,37.21)(0.14,-0.48){1}{\line(0,-1){0.48}}
\multiput(214.33,37.69)(0.16,-0.48){1}{\line(0,-1){0.48}}
\multiput(214.16,38.17)(0.17,-0.47){1}{\line(0,-1){0.47}}
\multiput(213.98,38.63)(0.09,-0.23){2}{\line(0,-1){0.23}}
\multiput(213.78,39.1)(0.1,-0.23){2}{\line(0,-1){0.23}}
\multiput(213.56,39.55)(0.11,-0.23){2}{\line(0,-1){0.23}}
\multiput(213.33,40)(0.11,-0.22){2}{\line(0,-1){0.22}}
\multiput(213.09,40.44)(0.12,-0.22){2}{\line(0,-1){0.22}}
\multiput(212.83,40.88)(0.13,-0.22){2}{\line(0,-1){0.22}}
\multiput(212.56,41.3)(0.13,-0.21){2}{\line(0,-1){0.21}}
\multiput(212.28,41.72)(0.14,-0.21){2}{\line(0,-1){0.21}}
\multiput(211.99,42.12)(0.15,-0.2){2}{\line(0,-1){0.2}}
\multiput(211.68,42.52)(0.1,-0.13){3}{\line(0,-1){0.13}}
\multiput(211.36,42.91)(0.11,-0.13){3}{\line(0,-1){0.13}}
\multiput(211.02,43.29)(0.11,-0.13){3}{\line(0,-1){0.13}}
\multiput(210.68,43.66)(0.11,-0.12){3}{\line(0,-1){0.12}}
\multiput(210.32,44.01)(0.12,-0.12){3}{\line(1,0){0.12}}
\multiput(209.95,44.36)(0.12,-0.11){3}{\line(1,0){0.12}}
\multiput(209.58,44.69)(0.13,-0.11){3}{\line(1,0){0.13}}
\multiput(209.19,45.01)(0.13,-0.11){3}{\line(1,0){0.13}}
\multiput(208.79,45.32)(0.13,-0.1){3}{\line(1,0){0.13}}
\multiput(208.38,45.61)(0.2,-0.15){2}{\line(1,0){0.2}}
\multiput(207.97,45.9)(0.21,-0.14){2}{\line(1,0){0.21}}
\multiput(207.54,46.17)(0.21,-0.13){2}{\line(1,0){0.21}}
\multiput(207.11,46.42)(0.22,-0.13){2}{\line(1,0){0.22}}
\multiput(206.67,46.67)(0.22,-0.12){2}{\line(1,0){0.22}}
\multiput(206.22,46.89)(0.22,-0.11){2}{\line(1,0){0.22}}
\multiput(205.76,47.11)(0.23,-0.11){2}{\line(1,0){0.23}}
\multiput(205.3,47.31)(0.23,-0.1){2}{\line(1,0){0.23}}
\multiput(204.83,47.49)(0.23,-0.09){2}{\line(1,0){0.23}}
\multiput(204.36,47.67)(0.47,-0.17){1}{\line(1,0){0.47}}
\multiput(203.88,47.82)(0.48,-0.16){1}{\line(1,0){0.48}}
\multiput(203.4,47.96)(0.48,-0.14){1}{\line(1,0){0.48}}
\multiput(202.91,48.09)(0.49,-0.13){1}{\line(1,0){0.49}}
\multiput(202.42,48.2)(0.49,-0.11){1}{\line(1,0){0.49}}
\multiput(201.92,48.29)(0.49,-0.09){1}{\line(1,0){0.49}}
\multiput(201.43,48.37)(0.5,-0.08){1}{\line(1,0){0.5}}
\multiput(200.93,48.43)(0.5,-0.06){1}{\line(1,0){0.5}}
\multiput(200.42,48.48)(0.5,-0.05){1}{\line(1,0){0.5}}
\multiput(199.92,48.51)(0.5,-0.03){1}{\line(1,0){0.5}}
\multiput(199.42,48.53)(0.5,-0.02){1}{\line(1,0){0.5}}
\put(198.91,48.53){\line(1,0){0.5}}
\multiput(198.41,48.51)(0.5,0.02){1}{\line(1,0){0.5}}
\multiput(197.91,48.48)(0.5,0.03){1}{\line(1,0){0.5}}
\multiput(197.41,48.43)(0.5,0.05){1}{\line(1,0){0.5}}
\multiput(196.91,48.37)(0.5,0.06){1}{\line(1,0){0.5}}
\multiput(196.41,48.29)(0.5,0.08){1}{\line(1,0){0.5}}
\multiput(195.92,48.2)(0.49,0.09){1}{\line(1,0){0.49}}
\multiput(195.42,48.09)(0.49,0.11){1}{\line(1,0){0.49}}
\multiput(194.94,47.96)(0.49,0.13){1}{\line(1,0){0.49}}
\multiput(194.45,47.82)(0.48,0.14){1}{\line(1,0){0.48}}
\multiput(193.97,47.67)(0.48,0.16){1}{\line(1,0){0.48}}
\multiput(193.5,47.49)(0.47,0.17){1}{\line(1,0){0.47}}
\multiput(193.03,47.31)(0.23,0.09){2}{\line(1,0){0.23}}
\multiput(192.57,47.11)(0.23,0.1){2}{\line(1,0){0.23}}
\multiput(192.11,46.89)(0.23,0.11){2}{\line(1,0){0.23}}
\multiput(191.67,46.67)(0.22,0.11){2}{\line(1,0){0.22}}
\multiput(191.22,46.42)(0.22,0.12){2}{\line(1,0){0.22}}
\multiput(190.79,46.17)(0.22,0.13){2}{\line(1,0){0.22}}
\multiput(190.37,45.9)(0.21,0.13){2}{\line(1,0){0.21}}
\multiput(189.95,45.61)(0.21,0.14){2}{\line(1,0){0.21}}
\multiput(189.54,45.32)(0.2,0.15){2}{\line(1,0){0.2}}
\multiput(189.14,45.01)(0.13,0.1){3}{\line(1,0){0.13}}
\multiput(188.76,44.69)(0.13,0.11){3}{\line(1,0){0.13}}
\multiput(188.38,44.36)(0.13,0.11){3}{\line(1,0){0.13}}
\multiput(188.01,44.01)(0.12,0.11){3}{\line(1,0){0.12}}
\multiput(187.66,43.66)(0.12,0.12){3}{\line(1,0){0.12}}
\multiput(187.31,43.29)(0.11,0.12){3}{\line(0,1){0.12}}
\multiput(186.98,42.91)(0.11,0.13){3}{\line(0,1){0.13}}
\multiput(186.66,42.52)(0.11,0.13){3}{\line(0,1){0.13}}
\multiput(186.35,42.12)(0.1,0.13){3}{\line(0,1){0.13}}
\multiput(186.05,41.72)(0.15,0.2){2}{\line(0,1){0.2}}
\multiput(185.77,41.3)(0.14,0.21){2}{\line(0,1){0.21}}
\multiput(185.5,40.88)(0.13,0.21){2}{\line(0,1){0.21}}
\multiput(185.24,40.44)(0.13,0.22){2}{\line(0,1){0.22}}
\multiput(185,40)(0.12,0.22){2}{\line(0,1){0.22}}
\multiput(184.77,39.55)(0.11,0.22){2}{\line(0,1){0.22}}
\multiput(184.56,39.1)(0.11,0.23){2}{\line(0,1){0.23}}
\multiput(184.36,38.63)(0.1,0.23){2}{\line(0,1){0.23}}
\multiput(184.17,38.17)(0.09,0.23){2}{\line(0,1){0.23}}
\multiput(184,37.69)(0.17,0.47){1}{\line(0,1){0.47}}
\multiput(183.85,37.21)(0.16,0.48){1}{\line(0,1){0.48}}
\multiput(183.71,36.73)(0.14,0.48){1}{\line(0,1){0.48}}
\multiput(183.58,36.24)(0.13,0.49){1}{\line(0,1){0.49}}
\multiput(183.47,35.75)(0.11,0.49){1}{\line(0,1){0.49}}
\multiput(183.38,35.26)(0.09,0.49){1}{\line(0,1){0.49}}
\multiput(183.3,34.76)(0.08,0.5){1}{\line(0,1){0.5}}
\multiput(183.23,34.26)(0.06,0.5){1}{\line(0,1){0.5}}
\multiput(183.19,33.76)(0.05,0.5){1}{\line(0,1){0.5}}
\multiput(183.15,33.26)(0.03,0.5){1}{\line(0,1){0.5}}
\multiput(183.14,32.75)(0.02,0.5){1}{\line(0,1){0.5}}
\put(183.14,32.25){\line(0,1){0.5}}
\multiput(183.14,32.25)(0.02,-0.5){1}{\line(0,-1){0.5}}
\multiput(183.15,31.74)(0.03,-0.5){1}{\line(0,-1){0.5}}
\multiput(183.19,31.24)(0.05,-0.5){1}{\line(0,-1){0.5}}
\multiput(183.23,30.74)(0.06,-0.5){1}{\line(0,-1){0.5}}
\multiput(183.3,30.24)(0.08,-0.5){1}{\line(0,-1){0.5}}
\multiput(183.38,29.74)(0.09,-0.49){1}{\line(0,-1){0.49}}
\multiput(183.47,29.25)(0.11,-0.49){1}{\line(0,-1){0.49}}
\multiput(183.58,28.76)(0.13,-0.49){1}{\line(0,-1){0.49}}
\multiput(183.71,28.27)(0.14,-0.48){1}{\line(0,-1){0.48}}
\multiput(183.85,27.79)(0.16,-0.48){1}{\line(0,-1){0.48}}
\multiput(184,27.31)(0.17,-0.47){1}{\line(0,-1){0.47}}
\multiput(184.17,26.83)(0.09,-0.23){2}{\line(0,-1){0.23}}
\multiput(184.36,26.37)(0.1,-0.23){2}{\line(0,-1){0.23}}
\multiput(184.56,25.9)(0.11,-0.23){2}{\line(0,-1){0.23}}
\multiput(184.77,25.45)(0.11,-0.22){2}{\line(0,-1){0.22}}
\multiput(185,25)(0.12,-0.22){2}{\line(0,-1){0.22}}
\multiput(185.24,24.56)(0.13,-0.22){2}{\line(0,-1){0.22}}
\multiput(185.5,24.12)(0.13,-0.21){2}{\line(0,-1){0.21}}
\multiput(185.77,23.7)(0.14,-0.21){2}{\line(0,-1){0.21}}
\multiput(186.05,23.28)(0.15,-0.2){2}{\line(0,-1){0.2}}
\multiput(186.35,22.88)(0.1,-0.13){3}{\line(0,-1){0.13}}
\multiput(186.66,22.48)(0.11,-0.13){3}{\line(0,-1){0.13}}
\multiput(186.98,22.09)(0.11,-0.13){3}{\line(0,-1){0.13}}
\multiput(187.31,21.71)(0.11,-0.12){3}{\line(0,-1){0.12}}
\multiput(187.66,21.34)(0.12,-0.12){3}{\line(1,0){0.12}}
\multiput(188.01,20.99)(0.12,-0.11){3}{\line(1,0){0.12}}
\multiput(188.38,20.64)(0.13,-0.11){3}{\line(1,0){0.13}}
\multiput(188.76,20.31)(0.13,-0.11){3}{\line(1,0){0.13}}
\multiput(189.14,19.99)(0.13,-0.1){3}{\line(1,0){0.13}}
\multiput(189.54,19.68)(0.2,-0.15){2}{\line(1,0){0.2}}
\multiput(189.95,19.39)(0.21,-0.14){2}{\line(1,0){0.21}}
\multiput(190.37,19.1)(0.21,-0.13){2}{\line(1,0){0.21}}
\multiput(190.79,18.83)(0.22,-0.13){2}{\line(1,0){0.22}}
\multiput(191.22,18.58)(0.22,-0.12){2}{\line(1,0){0.22}}
\multiput(191.67,18.33)(0.22,-0.11){2}{\line(1,0){0.22}}
\multiput(192.11,18.11)(0.23,-0.11){2}{\line(1,0){0.23}}
\multiput(192.57,17.89)(0.23,-0.1){2}{\line(1,0){0.23}}
\multiput(193.03,17.69)(0.23,-0.09){2}{\line(1,0){0.23}}
\multiput(193.5,17.51)(0.47,-0.17){1}{\line(1,0){0.47}}
\multiput(193.97,17.33)(0.48,-0.16){1}{\line(1,0){0.48}}
\multiput(194.45,17.18)(0.48,-0.14){1}{\line(1,0){0.48}}
\multiput(194.94,17.04)(0.49,-0.13){1}{\line(1,0){0.49}}
\multiput(195.42,16.91)(0.49,-0.11){1}{\line(1,0){0.49}}
\multiput(195.92,16.8)(0.49,-0.09){1}{\line(1,0){0.49}}
\multiput(196.41,16.71)(0.5,-0.08){1}{\line(1,0){0.5}}
\multiput(196.91,16.63)(0.5,-0.06){1}{\line(1,0){0.5}}
\multiput(197.41,16.57)(0.5,-0.05){1}{\line(1,0){0.5}}
\multiput(197.91,16.52)(0.5,-0.03){1}{\line(1,0){0.5}}
\multiput(198.41,16.49)(0.5,-0.02){1}{\line(1,0){0.5}}
\put(198.91,16.47){\line(1,0){0.5}}
\multiput(199.42,16.47)(0.5,0.02){1}{\line(1,0){0.5}}
\multiput(199.92,16.49)(0.5,0.03){1}{\line(1,0){0.5}}
\multiput(200.42,16.52)(0.5,0.05){1}{\line(1,0){0.5}}
\multiput(200.93,16.57)(0.5,0.06){1}{\line(1,0){0.5}}
\multiput(201.43,16.63)(0.5,0.08){1}{\line(1,0){0.5}}
\multiput(201.92,16.71)(0.49,0.09){1}{\line(1,0){0.49}}
\multiput(202.42,16.8)(0.49,0.11){1}{\line(1,0){0.49}}
\multiput(202.91,16.91)(0.49,0.13){1}{\line(1,0){0.49}}
\multiput(203.4,17.04)(0.48,0.14){1}{\line(1,0){0.48}}
\multiput(203.88,17.18)(0.48,0.16){1}{\line(1,0){0.48}}
\multiput(204.36,17.33)(0.47,0.17){1}{\line(1,0){0.47}}
\multiput(204.83,17.51)(0.23,0.09){2}{\line(1,0){0.23}}
\multiput(205.3,17.69)(0.23,0.1){2}{\line(1,0){0.23}}
\multiput(205.76,17.89)(0.23,0.11){2}{\line(1,0){0.23}}
\multiput(206.22,18.11)(0.22,0.11){2}{\line(1,0){0.22}}
\multiput(206.67,18.33)(0.22,0.12){2}{\line(1,0){0.22}}
\multiput(207.11,18.58)(0.22,0.13){2}{\line(1,0){0.22}}
\multiput(207.54,18.83)(0.21,0.13){2}{\line(1,0){0.21}}
\multiput(207.97,19.1)(0.21,0.14){2}{\line(1,0){0.21}}
\multiput(208.38,19.39)(0.2,0.15){2}{\line(1,0){0.2}}
\multiput(208.79,19.68)(0.13,0.1){3}{\line(1,0){0.13}}
\multiput(209.19,19.99)(0.13,0.11){3}{\line(1,0){0.13}}
\multiput(209.58,20.31)(0.13,0.11){3}{\line(1,0){0.13}}
\multiput(209.95,20.64)(0.12,0.11){3}{\line(1,0){0.12}}
\multiput(210.32,20.99)(0.12,0.12){3}{\line(1,0){0.12}}
\multiput(210.68,21.34)(0.11,0.12){3}{\line(0,1){0.12}}
\multiput(211.02,21.71)(0.11,0.13){3}{\line(0,1){0.13}}
\multiput(211.36,22.09)(0.11,0.13){3}{\line(0,1){0.13}}
\multiput(211.68,22.48)(0.1,0.13){3}{\line(0,1){0.13}}
\multiput(211.99,22.88)(0.15,0.2){2}{\line(0,1){0.2}}
\multiput(212.28,23.28)(0.14,0.21){2}{\line(0,1){0.21}}
\multiput(212.56,23.7)(0.13,0.21){2}{\line(0,1){0.21}}
\multiput(212.83,24.12)(0.13,0.22){2}{\line(0,1){0.22}}
\multiput(213.09,24.56)(0.12,0.22){2}{\line(0,1){0.22}}
\multiput(213.33,25)(0.11,0.22){2}{\line(0,1){0.22}}
\multiput(213.56,25.45)(0.11,0.23){2}{\line(0,1){0.23}}
\multiput(213.78,25.9)(0.1,0.23){2}{\line(0,1){0.23}}
\multiput(213.98,26.37)(0.09,0.23){2}{\line(0,1){0.23}}
\multiput(214.16,26.83)(0.17,0.47){1}{\line(0,1){0.47}}
\multiput(214.33,27.31)(0.16,0.48){1}{\line(0,1){0.48}}
\multiput(214.49,27.79)(0.14,0.48){1}{\line(0,1){0.48}}
\multiput(214.63,28.27)(0.13,0.49){1}{\line(0,1){0.49}}
\multiput(214.75,28.76)(0.11,0.49){1}{\line(0,1){0.49}}
\multiput(214.86,29.25)(0.09,0.49){1}{\line(0,1){0.49}}
\multiput(214.96,29.74)(0.08,0.5){1}{\line(0,1){0.5}}
\multiput(215.04,30.24)(0.06,0.5){1}{\line(0,1){0.5}}
\multiput(215.1,30.74)(0.05,0.5){1}{\line(0,1){0.5}}
\multiput(215.15,31.24)(0.03,0.5){1}{\line(0,1){0.5}}
\multiput(215.18,31.74)(0.02,0.5){1}{\line(0,1){0.5}}

\end{picture}

}

\begin{document}

\baselineskip 24pt

\begin{center}
{\Large \bf  Riding Gravity Away from Doomsday}

\end{center}

\vskip .6cm
\medskip

\vspace*{4.0ex}

\baselineskip=18pt

\centerline{\large \rm Ashoke Sen}

\vspace*{4.0ex}

\centerline{\large \it Harish-Chandra Research Institute}
\centerline{\large \it  Chhatnag Road, Jhusi,
Allahabad 211019, India}

\vspace*{1.0ex}
\centerline{\small E-mail:  sen@mri.ernet.in}

\vspace*{5.0ex}

\centerline{\bf Abstract} \bigskip  
The discovery that most of the energy density in the universe is stored in 
the form of dark energy has profound consequences for our future. In particular
our current limited understanding of quantum theory of gravity indicates that 
some time in the future our universe will undergo a phase transition that will
destroy us and everything else around us instantaneously. 
However the laws of gravity
also suggest a way out --  some
of our descendants could 
survive this catastrophe by riding gravity away from the danger. 
In this essay I describe the tale of
this escape from doomsday.

\bigskip

{\it Essay written for the Gravity Research Foundation 2015 Awards for Essays on Gravitation}



\vfill \eject

\baselineskip=18pt

\section{Doomsday}

The discovery 
of the accelerated rate of expansion of the present universe has had profound
influence on our current understanding of the universe\cite{9805201,9812133}. 
The best explanation of this to date is the
presence of a small positive cosmological constant in Einstein's equations.
According to our current knowledge, about 70\% of the energy density in the universe today comes from
the cosmological constant and the rest comes from ordinary matter and dark matter. 

\begin{figure}
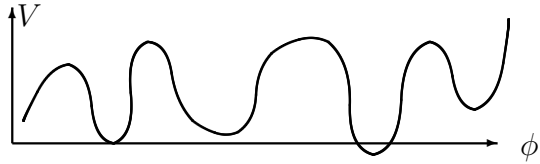


\begin{center}

\figa

\end{center}

\caption{\small This figure illustrates the 
landscape of phases in string theory. $\phi$ denotes a scalar
field and $V$ is the potential energy density.
\label{f1}}
\end{figure}

A plausible explanation of the
origin of the cosmological constant in string theory was provided in \cite{0301240,0004134,0105097}.
From the analysis of these papers and many subsequent papers the picture that has emerged is as
follows. The situation in string theory is analogous to (but far more complicated than) a field theory
of multiple scalar fields coupled to gravity, with the potential energy density 
being a complicated function of the fields with many local maxima, minima and saddle points. 
A one dimensional version of this has been shown in Fig.~\ref{f1}.
Each of the
local minima of the potential describes a phase of string theory, with the value of the potential energy density at
the minimum giving the value of the cosmological constant. The different phases 
have very different properties -- even the list of
`elementary particles' and the forces operating between them are very different. 
The hope is that someday
we shall find a minimum that describes exactly the kind of elementary particles and forces we see
in nature, but we are quite far from realizing this goal.

The description of different phases given above is classical. In quantum theory there are
fluctuations that keep driving the fields away from the local minimum. Occasionally
the fields inside a small region -- which we shall call the bubble -- 
would fluctuate to another minimum of the potential.
If the latter has lower potential energy density than the original minimum then the 
potential energy inside this bubble will be lowered 
if the bubble grows. 
However the surface tension of the bubble costs energy. 
Thus the net excess potential energy associated with
a bubble of radius $r$ can be written as
\be
-A \, r^3 + B \, r^2
\ee
where $A$ and $B$ are positive constants.
If $r>B/A$ this is negative. Such a bubble will begin to expand, with the excess
potential energy getting transfered to the wall as kinetic 
energy\cite{okun,stone,frampton,coleman1,coleman2,coleman}.
The wall
will begin to accelerate, with its speed eventually approaching the speed of light.
During this
expansion it will destroy everything that existed in the original phase
since even the elementary particles of the original phase do not 
exist in
the new phase in the interior of the bubble.\footnote{Gravity can make the
situation inside much worse\cite{coleman}.}

As I have already mentioned, we do not yet know our place in the vast landscape of phases of string theory.
But we do know that the minimum that describes us has a small but positive value of the potential energy
density. We also know that there are other minima of lower energy density
since there are many known
phases of string theory with zero and negative 
energy density. This has a sinister consequence: 
it shows that we cannot exist forever. Sooner or later in some part of the universe a 
bubble of another phase
of lower energy density will form which will subsequently expand 
and destroy us. Since the bubble wall expands at
the speed of light our death will be painless; we shall not even
know of the existence of the bubble before it hits us.
However this will definitely be the end of our species and the universe around us.

Even though we have used string theory to argue for the metastability of our
universe, there are various other independent arguments
which indicate that a space-time with positive
cosmological constant cannot exist 
forever\cite{0007146,0104180,0104181,0106109,0106113,0106209,0205316,0202163,0208013,0212209,0302219}.
String theory provides a concrete
realization of this phenomenon, but any other consistent quantum theory of
gravity will probably also lead to a similar conclusion.

How long do we have before such a calamity strikes us? 
Given that our universe has survived for 
$1.38\times 10^{10}$ years -- which is its current age -- 
it seems likely that
our half-life is not much smaller than $10^{10}$ years.\footnote{There
are different points of view on this issue, see {\it e.g.} \cite{bostron,0512204}.}
This means in particular that the probability that such a calamity will hit 
us during the
next one year is less that 1 part in $10^{10}$.
The actual probability may be
much smaller. 
Nevertheless if we wait long enough we are bound to encounter the doomsday
some time in the future unless other calamities have destroyed us by then.

\begin{figure}

\begin{center}


\vskip -1.9in

\epsfysize=11cm
\epsfbox{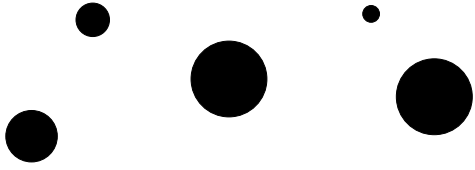}

\end{center}

\vskip -1.7in

\caption{\small This figure illustrates the formation and evolution 
of bubbles of destruction in our universe.
These bubbles are produced at random  at some constant rate, and once produced,
expand at the speed of light. At the same time, our universe expands exponentially,
with the distance between any two points roughly doubling 
every $10^{10}$ years. Unless the
bubble nucleation rate is too high the expansion of the universe wins, leaving us
with plenty of safe regions in between the bubbles. However since we do not know when 
and where a bubble might form, the only way to utilize these safe regions is to spread
out so that even when some of us are consumed by a bubble, others may survive.
\label{fd}}
\end{figure}

\section{The Escape}

The situation looks pretty bleak, particularly 
since we have no control on when and where such
a doomsday bubble may form!  Nevertheless there 
is a course of action that could save some of our descendants from this catastrophe. 
The essential idea is simple; we must spread out as fast as possible, 
establishing civilizations on different worlds
in different parts of the universe,  so that even if some of
us are hit by the catastrophe, the others may survive. 

\begin{figure}
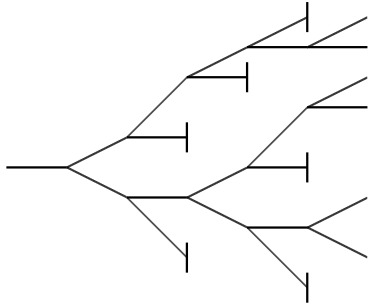


\begin{center}

\figb

\end{center}

\caption{\small This figure illustrates the multiplication of civilizations.
Here cosmic time -- measured by the logarithm of the 
inverse temperature of the microwave
background radiation -- runs from left to right 
and the bifurcations denote establishment of new civilizations.
Each civilization has a fixed probability per year of meeting its doomsday;
these doomsday events are marked by the vertical bars. 
While some of the civilizations meet their doomsday,
the others live on. If they multiply too slowly then it is likely that all the civilizations
will eventually meet their doomsday. On the other hand multiplying too fast does
not help since several civilizations within each other's horizon are likely to be annihilated by
the same bubble.
This figure has been inspired by a similar figure in \cite{1110.0496} 
where it was
used in a somewhat different context.  
\label{f5}}
\end{figure}

Since the bubble expands at the speed of light, it is not obvious that this strategy will work.
In fact, in Minkowski space-time this would require accelerating for ever.
But in a space-time like ours,
the
accelerated expansion of the universe helps by separating these different worlds 
so that eventually even
light sent from one of these civilizations cannot reach the other worlds. After this
any further communication between these worlds would be impossible -- they  
would have `gone outside each other's horizon'. In this situation 
a doomsday bubble hitting one of these
worlds will typically   be unable to reach the other worlds and destroy 
them\cite{guth}. Of course there may be
other bubbles hitting the other worlds at other times, but as long as we have sufficient number 
of them, some of the worlds may survive. This has been illustrated in Figs.~\ref{fd}, \ref{f5}.

In our universe the horizon size is of the order of $10^{10}$ light years. 
This means that
by sending out space-ships we can reach 
and establish civilizations on different worlds situated within a radius of about
$10^{10}$ light years from us today.\footnote{A person travelling at constant acceleration 
equal to the acceleration due to gravity on earth can
achieve this in her own life-time\cite{0509268}.}
As the universe expands these different worlds will get pulled apart from each
other. If they sit idle then eventually each of them will meet its doomsday.
To ensure the survival of our species, each of these civilizations must, in turn, spread out
and 
establish new civilizations. In order to optimize 
the survival probability we
have to ensure that at any given time, there is at least one civilization inside
a horizon size sphere -- a
sphere of radius $10^{10}$ light years.  Since the size of our universe doubles
approximately in every $10^{10}$ years, this would mean that the number of
civilizations must grow by a factor of eight in the same time.

Unfortunately this process cannot continue for ever.  Our resources are limited, determined by the 
fact that the total amount of matter at our disposal is what is contained inside the present horizon
radius of $10^{10}$ light years. This is a large amount of matter, but having to divide this into
the exponentially growing number of worlds will eventually 
reduce the total amount of matter available
to a given world to less than the critical value needed for establishing a 
civilization\cite{9902189}. 
Put another way, we would eventually run out of the worlds on which we can establish new
civilizations even if we manage to achieve a technological breakthrough and 
create new worlds by redistributing the available matter within our horizon.
Beyond this
point we shall no longer be able to establish new civilizations. Since each of the worlds has a finite
half-life, eventually each of them 
will encounter a doomsday bubble. Nevertheless we would have managed to prolong the
life of our species beyond what we would normally have by staying put in one place.

\section{Role of Fundamental Theory}

The cost of this endeavour is clearly going to be high. So 
it would make sense to follow this strategy only if the probability of the 
doomsday event is sufficiently high.
Precise information on this probability requires knowledge of the fundamental quantum theory
describing gravity and everything else. In the context of string theory this means that we need
to identify the correct minimum of the potential that describes the phase 
in which we
live and then compute the probability of decay of this phase by standard 
techniques. In searching for this minimum we can be guided by our knowledge of
low energy physics which is well described by the standard model. Another strong
constraint arises by demanding that the value of the quantum corrected potential
energy density 
at the minimum agrees with the observed value of the cosmological constant. 
Finding a minimum satisfying all these requirements is going to be
a herculean task, but given its bearing on the survival of our species, 
its perusal is worth the effort.

\bigskip

{\bf Acknowledgement:}
I wish to thank Rajesh Gopakumar, 
Dileep Jatkar, Anshuman Maharana and 
Sumathi Rao for their comments on the manuscript.
This work  was
supported in part by the 
DAE project 12-R\&D-HRI-5.02-0303 and J. C. Bose fellowship of 
the Department of Science and Technology, India.

\end{document}